\documentclass[a4paper,11pt]{article}

\pdfoutput=1
\usepackage{jheppub} 

\usepackage[T1]{fontenc} 

\usepackage{Settings}
\usepackage{tikz}
\usepackage{slashed}

\usepackage[font=small,labelfont=bf,width=0.85\textwidth]{caption}
\usepackage{subcaption}

\makeatletter
\newlength{\apb@width}
\newcommand{\autoparbox}[2][c]{\settowidth{\apb@width}{#2}\parbox[#1]{\apb@width}{#2}}

\makeatother

\renewcommand{\vec}[1]{\mathbf{#1}}

\makeatletter
\def\mr@ignsp#1 {\ifx\:#1\@empty\else #1\expandafter\mr@ignsp\fi}%
\newcommand{\multiref}[1]{\begingroup
	\xdef\mr@no@sparg{\expandafter\mr@ignsp#1 \: }%
	\def\mr@comma{}%
	\@for\mr@refs:=\mr@no@sparg\do{\mr@comma\def\mr@comma{,}\ref{\mr@refs}}%
	\endgroup}
\makeatother

\newcommand{\hypref}[2]{\ifx\href\asklfhas #2\else\href{#1}{#2}\fi}

\renewcommand{\eqref}[1]{(\multiref{#1})}

\usepackage{blindtext}%
\usepackage{etoolbox}%
\title{\boldmath Light-ray operators, detectors and gravitational event shapes}

\author[a]{Riccardo Gonzo} \author[b]{, Andrzej Pokraka} 

\affiliation[a]{School of Mathematics \& Hamilton Mathematics Institute,\\
	 Trinity College Dublin, Ireland}
\affiliation[b]{Department of Physics, McGill University,\\
	 3600 Rue University, Montr{\'e}al, QC, Canada H3A 2T8}
	 
\emailAdd{gonzo@maths.tcd.ie}
\emailAdd{andrzej@physics.mcgill.ca}

\abstract{Light-ray operators naturally arise from integrating Einstein equations at null infinity along the light-cone time. We associate light-ray operators to physical detectors on the celestial sphere and we provide explicit expressions in perturbation theory for their hard modes using the steepest descent technique. We then study their algebra in generic 4-dimensional QFTs of massless particles with integer spin, comparing with complexified Cordova-Shao algebra. For the case of gravity, the Bondi news squared term provides an extension of the ANEC operator at infinity to a shear-inclusive ANEC, which as a quantum operator gives the energy of all quanta of radiation in a particular direction on the sky. We finally provide a direct connection of the action of the shear-inclusive ANEC with detector event shapes and we study infrared-safe gravitational wave event shapes produced in the scattering of massive compact objects, computing the energy flux at infinity in the classical limit at leading order in the soft expansion.}

\begin{document} 

\maketitle
\flushbottom

\section{Introduction}
\label{sec:intro}

Event shapes or weighted cross-sections describe the distribution of outgoing particles and their properties such as charge or energy-momentum in a particular direction as determined by a detector. They are important tools for analyzing jets produced in collider experiments   \cite{Basham:1977iq,Basham:1978bw,Basham:1978zq}. In the '90s, Korchemsky and Sterman\footnote{See   \cite{Ore:1979ry,Sveshnikov:1995vi} for earlier approaches.} discovered that such event shapes can be equivalently described by the expectation value of some non-local operators -- line integrals of local conserved currents -- inserted at different locations on the celestial sphere   \cite{Korchemsky:1997sy,Korchemsky:1999kt}. While these objects are well-known in perturbative QCD, where they are used to study hadron scattering at high energies   \cite{Mueller:1998fv}, it is only recently that Maldacena and Hofman initiated a systematic investigation of such operators in conformal field theory  \citep{Hofman:2008ar}. It is now well understood how to extract event shapes from OPE data and symmetries in conformal field theories (especially for $\mathcal{N}=4$ SYM)   \cite{Belitsky:2013bja,Belitsky:2013xxa}. Moreover, the connection between energy event shapes and the ANEC operator establishes bounds on the \textit{a} and \textit{c} coefficients characterizing the conformal anomaly   \cite{Hofman:2008ar,Hofman:2016awc,Belin:2019mnx}. Of course, such developments led to increased interest in CFT light-ray operators which culminated in the systematic analysis of   \cite{Kravchuk:2018htv,Kologlu:2019mfz,Chang:2020qpj,Belin:2020lsr}. 

We can also study the correlation between event shapes by considering more than one detector (ie., multiple insertions of the corresponding non-local operators). Of particular interest is the so called energy-energy correlator, which is an infrared-finite observable that can be measured in collider experiments and used to provide a measure of the QCD coupling constant. The leading order QCD prediction for energy-energy correlators was computed in the late '70s   \cite{Basham:1978bw,Basham:1978zq}. More recently, analytical expressions at NLO   \cite{Dixon:2018qgp,Luo:2019nig} as well as numerical expressions at NNLO   \cite{DelDuca:2016ily} have become available. In $\mathcal{N}=4$ SYM, there are even analytic results at NNLO   \cite{Henn:2019gkr}. 

The recent discovery of gravitational waves  \cite{Abbott:2016blz,TheLIGOScientific:2017qsa} started a new era in multi-messenger astronomy.  A powerful technique to construct bank of template waveforms for the detectors is the effective one-body (EOB) map  \cite{Taracchini:2013rva,Bohe:2016gbl} developed in a seminal paper by A. Buonanno and T. Damour  \cite{Buonanno:1998gg}, which combines both analytical and numerical (when combined with numerical relativity (EOB-NR)) data for the description of the bound state problem in general relativity. It was pointed out by Damour   \cite{Damour:2016gwp,Damour:2017zjx} that the scattering problem and its post-Minkowskian expansion contains useful information which can help to construct the effective one-body hamiltonian for the inspiral phase. More recently the relation between bound and unbound orbits has been better clarified, at least in the conservative regime  \cite{Kalin:2019rwq,Kalin:2019inp,Bini:2020flp}.

Scattering amplitudes and on-shell techniques have proven to be very useful in order to study the conservative regime for compact binaries. In an EFT approach  \cite{Goldberger:2004jt} such objects are treated as point particles with possible additional finite size effect operators. Taking the classical limit from quantum scattering amplitudes requires to understand which type of coupling gives the classical spin dynamics  \cite{Guevara:2018wpp,Arkani-Hamed:2019ymq,Guevara:2019fsj}, which was referred as ``minimal coupling'' in the literature  \cite{Arkani-Hamed:2017jhn,Chung:2018kqs}. With the matching with a suitable EFT potential  \cite{Cheung:2018wkq} and combining several techniques like double copy and unitarity, it was possible to extract for the first time the conservative binary hamiltonian\footnote{See also \cite{Bjerrum-Bohr:2019kec,Cristofoli:2019neg}.} at $G^3$ order for the spinless case  \cite{Bern:2019nnu,Bern:2019crd} and at $G^2$ order for the spinning case  \cite{Bern:2020buy}. This result has passed many non-trivial checks at the quantum   \cite{Cheung:2020gyp} and at the classical level  \cite{Kalin:2020fhe,Kalin:2020mvi,Damour:2020tta}. There has been some attempts to study radiation in this context both using classical worldline calculations   \cite{Goldberger:2016iau,Goldberger:2017ogt,Shen:2018ebu,delaCruz:2020bbn,Plefka:2019hmz,Bonocore:2020xuj,Mogull:2020sak} and double copy techniques scattering amplitudes in the classical limit  \cite{Luna:2017dtq,Bautista:2019tdr}. In addition, several classical observables like the classical impulse  \cite{Kosower:2018adc}, the radiation kernel  \cite{Kosower:2018adc} and the classical spin kick  \cite{Maybee:2019jus,Guevara:2019fsj} were defined in full generality from from the classical limit of matrix elements of appropriate quantum operators. 

Here we would like to understand how the energy carried away by gravitational waves, which is collected by a detector localized in a direction $\hat{\vec{n}}$, is described at the quantum level. We will see that the corresponding operator is related to the Isaacson effective stress tensor  \cite{Isaacson:1967zz,Isaacson:1968zza} by the integration over the retarded working time of the detector, and its action can be computed explicitly using standard techniques in the asymptotic expansion.  Using the Bondi gauge framework for asymptotically flat spacetimes, we will show that the energy flow operator can be also written in terms of the Bondi news squared term and this provides an extension of ANEC operator at null infinity to a shear inclusive ANEC. With our detector interpretation, the latter naturally provides the sum of energies of all massless quanta (i.e. radiation) emitted in a direction $\hat{\vec{n}}$ when acting on on-shell states. One can then generalize the standard notion of event shapes used in QCD to the gravitational case, which we will study in detail in the soft kinematic regime by taking the classical limit and summing explicitly over all graviton contributions. Indeed in the very low-energy regime, a single coherent state provides the exact quantum state of radiation and taking the limit of the large number of quanta reproduce known classical limit results for the energy flux  \cite{Smarr:1977fy,GellMann:1954kc,Low:1958sn,Weinberg:1965nx}. A systematic study of the low-energy gravitational radiation produced in the scattering of compact objects up to sub-subleading order in the soft expansion has already been performed in the literature, both in the quantum and in the classical context  \cite{Laddha:2018rle,Laddha:2018myi,Sahoo:2018lxl,Laddha:2019yaj,Saha:2019tub,Addazi:2019mjh,Ciafaloni:2018uwe,A:2020lub}. We will also study the correlation between different gravitational energy event shapes, especially in relation to the expected classical soft factorization. 

Light-ray operators constructed from the components of the stress tensor are naturally related to the global Poincar\'e charges by turning the line integral in the light-ray definition into an integral over all of space. We will restrict our analysis to the subset of light-ray operators obtained by integrating components of the stress tensor over a null-sheet. These light-ray operators were recently studied in the context of unitary CFTs by Cordova and Shao \cite{Cordova:2018ygx} and shown, via symmetry arguments, to form a closed algebra. While it is not immediately clear that the same light-ray operators will form a closed algebra in a generic QFT, the underlying universality of the stress tensor algebra \cite{Deser:1967zzf} hints at this possibility. The stress tensor algebra was also reconsidered  more recently \cite{Huang:2019fog,Huang:2020ycs} to derive a ``universal'' effective light-cone algebra for CFTs in $d > 2$. Moreover, there is an intriguing relation with the BMS algebra \cite{Bondi:1962px,Sachs:1962wk} -- and its extended version  \cite{Barnich:2009se,Barnich:2011ct,Barnich:2011mi,Campiglia:2014yka,Strominger:2017zoo} -- with light-ray operators  \cite{Cordova:2018ygx} which is still worth exploring for light-ray placed at null infinity. Motivated by that, we will define a new system of light-ray operators for linearized gravity, which will naturally combine with the standard stress tensor definition in scalar and gauge theories providing a unified treatment of all massless particles at $\mathcal{I}^{\pm}$.  We will then compute in our approach the associated algebra of light-ray operators in $3+1$-dimensional QFTs for massless particles of integer spin and compare with \cite{Cordova:2018ygx}.

\paragraph{Preliminaries}

Let's consider the process of electron-positron annihilation in QCD: for a generic outcome of the scattering process
\begin{align}
e^+ \, e^- \to X ,
\label{eqn:scattering_e+e-}
\end{align}
we would like to understand the properties of the emitted particles as captured by one (or many) physical detectors located at spatial infinity. To make contact with the structure at null infinity, it is convenient to work with flat null coordinates for Minkowski spacetime
\begin{align}
\mathrm{d} s^2 = \mathrm{d} u \mathrm{d} r -  r^2  \mathrm{d} z \mathrm{d} \bar{z} 
\label{eqn:Bondi_flat}
\end{align}
so that the null boundaries are located at $r \to +\infty$ while keeping $(u,z,\bar{z})$ fixed. On this hypersurface $(z,\bar{z})$ are \emph{stereographic coordinates} on the celestial sphere. This set of coordinates correspond to wrapping up the transverse spatial coordinates of the light-sheet at infinity 
\begin{align}
\mathrm{d} s^2 = \mathrm{d} x^- \mathrm{d} x^+ - \mathrm{d} x^1 \mathrm{d} x^1 - \mathrm{d} x^2 \mathrm{d} x^2
\end{align}
via the transformation
\begin{align}
x^- &\to u + r z \bar{z} \qquad x^+ \to r \qquad x^1 \to \frac{1}{2} r (z+\bar{z}) \qquad x^2 \to -\frac{1}{2} i r (z-\bar{z}) .
\label{eqn:light_sheet_wrapping}
\end{align}

Instead of considering the total cross section for the process \eqref{eqn:scattering_e+e-}
\begin{align}
\sigma_{\text{tot}}(p_1,p_2) :=\sum_X (2 \pi)^4 \delta^4 (p_1 + p_2 - k_X) |\mathcal{M}_{e^+ \, e^- \to X}|^2 ,
\end{align}
where $p_1$ and $p_2$ are the incoming momenta of $e^+$ and $e^-$ and $k_X$ is the sum of the outgoing momenta, we introduce a less inclusive observable by defining a weight $w(X)$ such that
\begin{align}
\sigma_W(p_1,p_2) &:= \frac{1}{\sigma_{\text{tot}}(p_1,p_2)} \sum_X (2 \pi)^4 \delta^4 (p_1 + p_2 - k_X) w(X) |\mathcal{M}_{e^+ \, e^- \to X}|^2 \nonumber \\
&= \frac{1}{\sigma_{\text{tot}}(p_1,p_2)} \sum_X \bra{\vec{p}_1 \vec{p}_2} S^{\dagger} \ket{\vec{X}} w(X) \bra{\vec{X}} S \ket{\vec{p}_1 \vec{p}_2} ,
\end{align}
which is the so-called ``weighted cross section''. Different choices of the weight factors $w(X)$ can tell us about different properties of the distribution of outgoing particles, like the flow of quantum number or of energy and momentum in a particular direction on the celestial sphere. 

Of particular interest is the to the choice of $w(X)$ corresponding to the energy flow. Given an external on-shell state $\ket{\vec{X}}=\ket{\vec{k}_1 \dots \vec{k}_n}$ of $n$ massless particles with $k_X^{\mu} = \sum_{i=1}^n k_i^{\mu}$, we define
\begin{align}
\label{eq:ANECWeight}
w_{\mathcal{\tilde{E}}}(X)(k_1,...,k_n) := \sum_{i=1}^n E_i \delta^2(\Omega_{\vec{\hat{k}}_i} - \Omega_{\hat{\vec{n}}}) ,
\end{align}
where $k_i^{\mu} = (E_i,\vec{k}_i)$ is the 4-momentum of the individual particles and $\Omega_{\vec{\hat{k}}_i} = \frac{\vec{k}_i}{|\vec{k}_i|}$ is the solid angle in the direction of $\vec{k}_i$  \cite{Belitsky:2013bja,Belitsky:2013xxa}. The weighted cross section measures the distribution of energy in the final state that flows in the direction of $\hat{\vec{n}}$, and as such we can interpret $w_{\mathcal{E}}(X)$ as an eigenvalue of the average null energy operator (ANEC) at \emph{spatial infinity}
\begin{align}
\mathcal{\tilde{E}}(\hat{\vec{n}}) =  \int_{0}^{+\infty} \mathrm{d} t \, \lim_{r \to +\infty} r^2 \, \hat{n}^i T_{0 i}(t, r \hat{\vec{n}}) ,
\label{eqn:ANEC}
\end{align}
where the energy momentum tensor is always understood to be normal ordered and $t$ is the physical working time of the detector. Looking at the energy flow operator \eqref{eqn:ANEC} it is easy to generalize it to a momentum flow operator (which is well-known in the jet literature  \cite{Ellis:2010rwa,Bauer:2008dt})
\begin{align}
\mathcal{\tilde{P}}_{\mu}(\hat{\vec{n}}) =  \int_{0}^{+\infty} \mathrm{d} t \, \lim_{r \to +\infty} r^2 \, \hat{n}^i T_{\mu i}(t, r \hat{\vec{n}}) ,
\label{eqn:momentum_flow2}
\end{align}
whose action on a state gives the linear momentum flowing the particular direction $\hat{\vec{n}}$
\begin{align}
w_{\mathcal{\tilde{P}}_j}(X)(k_1,...,k_n) := \sum_{i=1}^n (\vec{k}_j)_i \delta^2(\Omega_{\vec{\hat{k}}_i} - \Omega_{\hat{\vec{n}}}) .
\end{align}

One can define also the expectation value of the ANEC operator at infinity, namely the 1-point energy event shape
\begin{align}
\langle \mathcal{\tilde{E}}(\hat{\vec{n}}) \rangle := \frac{1}{\sigma_{\text{tot}}(p_1,p_2)} \bra{\vec{p}_1 \vec{p}_2} S^{\dagger} \mathcal{\tilde{E}}(\hat{\vec{n}}) S \ket{\vec{p}_1 \vec{p}_2} ,
\label{eqn:1_pt_ANEC}
\end{align}
where it is always understood that $\mathcal{\tilde{E}}(\hat{\vec{n}})$ acts on the set of outgoing states $\ket{\vec{X}}$ inserted via the completeness relation, as usual in the Schwinger-Keldysh formalism.

\begin{figure}[h!]
\centering
  \centering
  \includegraphics[width=0.60\textwidth]{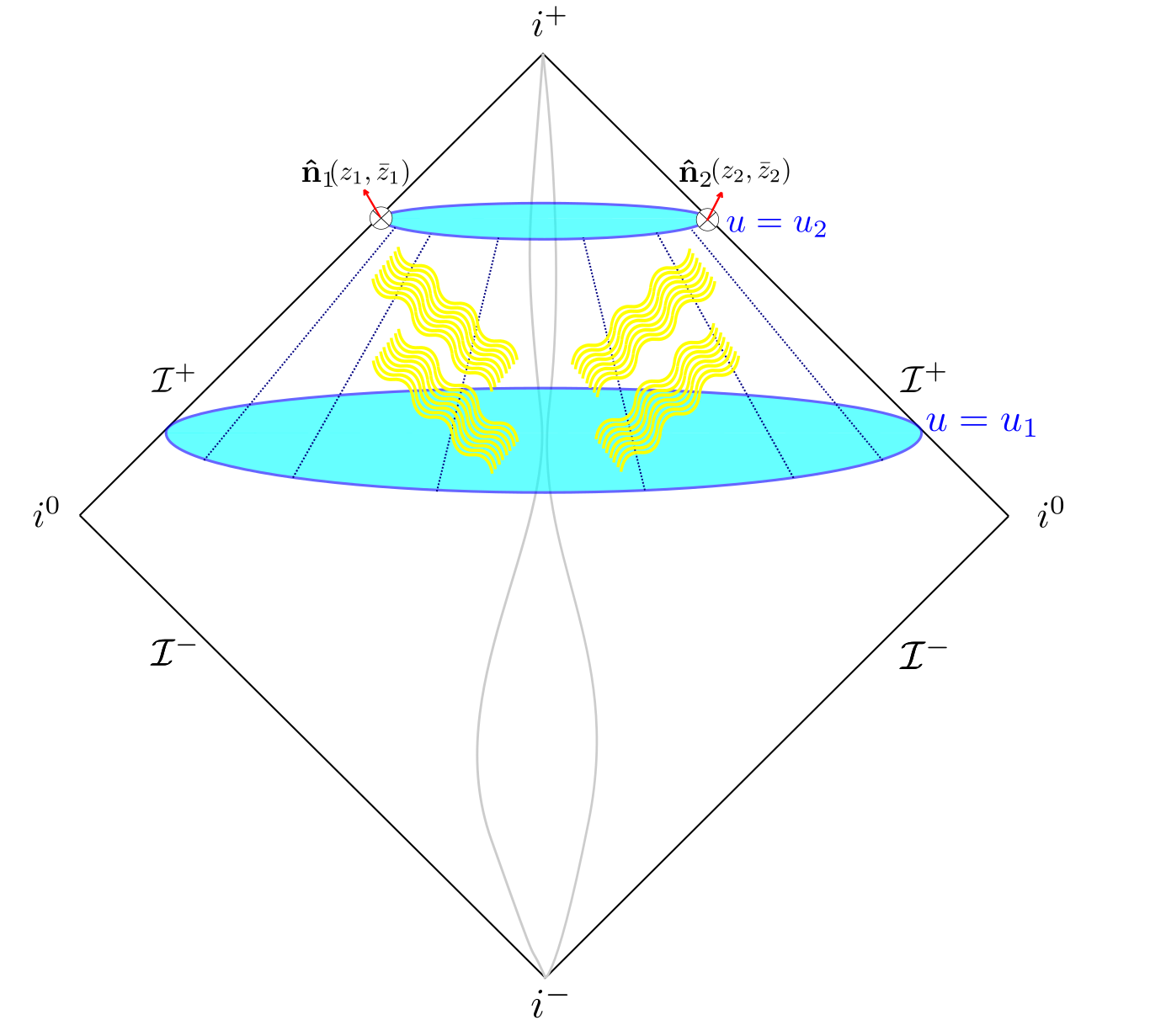}
  \caption{The radiation due to the scattering of two massive objects is captured by two detectors located at spatial infinity in the directions $\hat{\vec{n}}_1(z_1,\bar{z}_1)$ and $\hat{\vec{n}}_2(z_2,\bar{z}_2)$.}
\end{figure}

In the case of two (or more) detectors, as previously mentioned, we can also define the energy-energy correlator which is related to the 2-pt \emph{Wightman} function of two ANEC operators at infinity \eqref{eqn:ANEC} inserted at different points on the celestial sphere
\begin{align}
\langle \mathcal{\tilde{E}}(\hat{\vec{n}}) \mathcal{\tilde{E}}(\hat{\vec{n}}^{\prime}) \rangle := \frac{1}{\sigma_{\text{tot}}(p_1,p_2)} \bra{\vec{p}_1 \vec{p}_2} S^{\dagger} \mathcal{\tilde{E}}(\hat{\vec{n}}) \mathcal{\tilde{E}}(\hat{\vec{n}}^{\prime}) S \ket{\vec{p}_1 \vec{p}_2} .
\label{eqn:2_pt_ANEC}
\end{align}
It is quite interesting to analyze this 2-pt correlator for the gravitational radiation in the classical limit, where in the most general case a superposition of coherent states represents the radiation at the quantum level.

In flat null coordinates the momentum flow operator can be written directly in terms of the energy flow
\begin{align}
\mathcal{P}^{\mu}(\hat{\vec{n}}) =  n^{\mu} \mathcal{E}(\hat{\vec{n}}) = n^{\mu} \int_{-\infty}^{+\infty} \mathrm{d} u \, \lim_{r \to \infty} r^2 T_{u u}(u,r,z_{\hat{\vec{n}}},\bar{z}_{\hat{\vec{n}}}) ,
\label{eqn:flat_null_E_P}
\end{align}
where $u$ is the retarded time and identify $\hat{\vec{n}}$ with the related coordinates on the celestial sphere
\begin{align}
\hat{\vec{n}} \Leftrightarrow (z_{\hat{\vec{n}}},\bar{z}_{\hat{\vec{n}}}) .
\end{align}
Here, we emphasize that we use $\mathcal{E}(\hat{\vec{n}})$ in place of $\mathcal{\tilde{E}}(\hat{\vec{n}})$ for technical convenience. One can work in Bondi gauge where the detector would register the energy flow as given by $\mathcal{\tilde{E}}(\hat{\vec{n}})$ (i.e. the standard local energy factor). We will do this later on for gravitational event shapes.

\section{Light-ray operators in QFT}
\label{sec:def}

In this section, we define the family of light-ray operators under consideration and make contact with their physical significance and in particular with the corresponding surface charges of null-sheet symmetry generators. Light-ray operators depend implicitly on a choice of null-sheet and we choose future null infinity. 

In flat-null coordinates, our family of light-ray operators is defined by
\begin{align}
\mathcal{E}(\hat{\vec{n}}) &= \int_{-\infty}^{+\infty} \mathrm{d} u \, \lim_{r \to \infty} r^2 T_{u u}(u,r,z_{\hat{\vec{n}}},\bar{z}_{\hat{\vec{n}}}) ,
\nonumber \\
\mathcal{K}(\hat{\vec{n}}) &= \int_{-\infty}^{+\infty} \mathrm{d} u \, u \lim_{r \to \infty} r^2 T_{u u}(u,r,z_{\hat{\vec{n}}},\bar{z}_{\hat{\vec{n}}}) ,
\nonumber \\
\mathcal{N}_z(\hat{\vec{n}}) &=  \int_{-\infty}^{+\infty} \mathrm{d} u \, \lim_{r \to \infty}  r^2 T_{u z}(u,r,z_{\hat{\vec{n}}},\bar{z}_{\hat{\vec{n}}}) ,
\nonumber \\
\mathcal{N}_{\bar{z}}(\hat{\vec{n}}) &=  \int_{-\infty}^{+\infty} \mathrm{d} u \, \lim_{r \to \infty} r^2 T_{u \bar{z}}(u,r,z_{\hat{\vec{n}}},\bar{z}_{\hat{\vec{n}}}) .
\end{align}
We will collectively denote these operators by $L(\hat{\vec{n}}) =\{\mathcal{E}(\hat{\vec{n}}),\mathcal{K}(\hat{\vec{n}}),\mathcal{N}_{z}(\hat{\vec{n}}),\mathcal{N}_{\bar{z}}(\hat{\vec{n}})\}$.

This family of light-ray operators can be understood as \emph{surface densities}\footnote{There are various subtleties in such definition as we will discuss explicitly here. Technically surface densities are defined up to boundary terms that do not change the total charge. Thus, only \emph{the equivalence class} of a light-ray operator is uniquely defined. Moreover we are interested in the standard (hard) charge coming from contracting the stress tensor $T_{\mu \nu}$ with a Killing vector $\xi^{\mu}$, which might differ from the isometry charge in the covariant phase space formulation.} for the corresponding conserved charges of the future null boundary of Minkowski spacetime. To see this, consider an affine transformation of the null surface generator   \cite{Wall:2011hj}
\begin{align}
\delta u = A + B \, u ,
\end{align}
where $A$ and $B$ are arbitrary constants. Since $\lim_{r \to \infty} r^2 T_{u u}(u,r,z,\bar{z})$ corresponds to the generator of null diffeomorphisms, we can integrate it to obtain either the generator of null translation $\delta u = A $
\begin{align}
\frac{1}{2} \int_{-\infty}^{+\infty} \mathrm{d} u \int \mathrm{d}^2 z \, \lim_{r \to \infty} r^2 T_{u u}(u,r,z,\bar{z})
\end{align}
or of dilatations $\delta u = B \, u$
\begin{align}
\frac{1}{2} \int_{-\infty}^{+\infty} \mathrm{d} u \int \mathrm{d}^2 z \, u \lim_{r \to \infty} r^2 T_{u u}(u,r,z,\bar{z}) ,
\end{align}
which is an equivalent definition of $\int \mathrm{d}^2 z \, \mathcal{K}(z,\bar{z})$. The latter is called ``boost mass'' $M_{\text{boost}}$ in the literature and it is usually defined for null-like horizons\footnote{More generically, choosing a spatial slice of the null surface $u^0 = \Lambda(z,\bar{z})$, a canonical ``boost energy'' can be defined \cite{Wall:2011hj}
\begin{align}
\frac{1}{2} \int \mathrm{d}^2 z \int_{\Lambda(z,\bar{z})}^{+\infty} \mathrm{d} u \, (u - \Lambda(z,\bar{z})) \lim_{r \to \infty} r^2  T_{u u}(u,r,z,\bar{z}) ,
\end{align}
which turns out to be connected with the notion of ``area operator'', providing thus a link to the concept of generalized entropy and to the generalized second law.} using the boost Killing vector of a causal wedge in Minkowski spacetime \cite{Jacobson:1995ab,Wall:2010cj,Wall:2011hj}. In a CFT, this provides a direct connection to the concept of modular hamiltonian for null-like horizons   \cite{Casini:2017roe}. While for empty Minkowski spacetime the boost mass is exactly zero, for more general non-asymptotically flat spacetimes it might be relevant for a proper formulation of the first law of black hole thermodynamics \cite{Dutta:2005iy}. In particular, it turns out that while the ADM mass $M_{\text{ADM}}$ measures the total ``monopole'' distribution of the matter in the spacetime, the boost mass $M_{\text{boost}}$ measures the total ``dipole'' moment of the mass distribution at infinity  \cite{Dutta:2005iy}. 

Since we are only considering QFTs on a flat Minkowski background, we will not delve deeper into these topics. However, we will find the monopole versus dipole distinction relevant for understanding the action of $\mathcal{K}(\hat{\vec{n}})$ on on-shell states. From now on, we will call $\mathcal{K}(\hat{\vec{n}})$ the \emph{boost energy (surface) density}. In the same spirit, the ANEC at infinity can be called \emph{null energy (surface) density}. 

Following a similar line of reasoning one finds that $\mathcal{N}_z(\hat{\vec{n}})$ and $\mathcal{N}_{\bar{z}}(\hat{\vec{n}})$ are associated to the local vector fields $\partial_z$ and $\partial_{\bar{z}}$, i.e. with the angular momentum flux charge
\begin{align}
\frac{1}{2} \int_{-\infty}^{+\infty} \mathrm{d} u \int \mathrm{d}^2 z \, \lim_{r \to \infty} r^2 T_{u z}(u,r,z,\bar{z}) \,, \qquad \frac{1}{2} \int_{-\infty}^{+\infty} \mathrm{d} u \int \mathrm{d}^2 z \, \lim_{r \to \infty} r^2 T_{u \bar{z}}(u,r,z,\bar{z}) .
\end{align}
Therefore $\mathcal{N}_z(\hat{\vec{n}})$ and $\mathcal{N}_{\bar{z}}(\hat{\vec{n}})$ can be thought as \emph{angular momentum flux (surface) densities}. More generally, this family of non-local light-ray operators appear naturally in Einstein equations solved near null infinity and in particular in (BMS and Poincar\'e) balance flux laws \cite{Compere:2019gft}, where they represent the radiative contribution to the fluxes.

Here we interpret physically the insertion of our light-ray operators as an insertion of a physical detector on the boundary of Minkowski spacetime. We will be interested in how these operators act on on-shell states and we will then compute their expectation values in perturbation theory, as in the standard event shapes literature. 

\section{The Spin-0 light-ray operators}

Here, we consider a theory of self-interacting massless scalars with a potential $Q(\phi)$ that is polynomial in the fields and without derivative interactions.  We derive explicit expressions for our family of light-ray operators in terms of the scalar creation and annihilation operators.

The Hilbert stress tensor $T^{\text{matter}}_{\mu \nu} = - \frac{2}{\sqrt{|g|}} \frac{\delta S^{\text{matter}}}{\delta g^{\mu \nu}}$ for our theory of scalars is simply 
\begin{align}
T_{\mu \nu}^{\text{scalar}} = (\partial_{\mu} \phi)(\partial_{\nu} \phi) -\eta_{\mu \nu} \left( \frac{1}{2} \partial_{\alpha} \phi \partial^{\alpha} \phi - Q(\phi) \right) .
\end{align}
The scalar stress tensor components which are relevant for the light-ray operators are\footnote{One can easily work in a more general setting with derivative interactions. However, such terms will drop off due to their scaling in $\frac{1}{r}$ in the limit that the null-sheet is pushed out to null-infinity. Instead, we prefer the interaction terms drop off directly from the structure of stress tensor components since this more closely resembles scalar light-ray operators in the bulk (see appendix C).}
\begin{align}
T^{\text{scalar}}_{u u}(x) {=} (\partial_u \phi) (\partial_u \phi), \qquad
T^{\text{scalar}}_{u z}(x) {=} (\partial_u \phi) (\partial_z \phi), \qquad 
T^{\text{scalar}}_{u \bar{z}}(x) {=} (\partial_u \phi) (\partial_{\bar{z}} \phi). 
\end{align}
The next step is to substitute the leading term in the large $r$ expansion of the scalar field into the expressions above. However, to do this, we need to impose some boundary conditions on the field.

Since we require the energy flux across $\mathcal{I}^+$ to be finite, we impose the following falloff condition for the scalar field\footnote{Soft scalar modes are beyond the scope of this work, so we do not allow the scalar field asymptotic expansion to have a constant piece around $\mathcal{I}^{+}_{\pm}$.}
\begin{align}
\phi (u,r,z,\bar{z}) &= \frac{\phi^{(0)} (u,z,\bar{z})}{r} + \mathcal{O}\left(\frac{1}{r}\right) , \qquad
\lim_{u \to \pm \infty} \phi^{(0)} (u,z,\bar{z}) = 0 .
\end{align}
Far away from interactions we can safely use the free field mode expansion
\begin{align}
\phi(x) = \int \widetilde{\mathrm{d}^{3} p} \left[b(\vec{p}) e^{-i p \cdot x} + b^{\dagger}(\vec{p}) e^{i p \cdot x} \right] 
\end{align}
and evaluate the field with the large $r$ saddle-point estimate. Physically, $r$ is the distance from the detector to the region where the interaction is localized and it should be much larger than the inverse of the smallest momentum scale that appears in the S-matrix \cite{Bauer:2008dt}.
Note that in our coordinate conventions the canonical commutators read
\begin{align}
	[b(\vec{p}),b^\dagger(\vec{q})] = \frac{4(2\pi)^3}{\omega_p}\delta(\omega_p-\omega_q) 
	\delta^2(z_{\hat{\vec{p}}}-z_{\hat{\vec{q}}}) .
\end{align}
The large $r$ limit of such a Fourier integral is controlled by the exponents of the exponentials  
\begin{align}
f(z,w) := i p \cdot x = i \frac{\omega_p u}{2} + i \frac{\omega_p r}{2} |z - w|^2 ,
\end{align}
where we use
\begin{align}
x^{\mu} &= \frac{r}{2} \left(1 + z \bar{z} + \frac{u}{r}, z +\bar{z},-i (z -\bar{z}),1 - z \bar{z} - \frac{u}{r}\right) , \nonumber \\
p^{\mu} &= \frac{\omega_p}{2} \left(1 + w \bar{w}, w +\bar{w},-i (w -\bar{w}),1 - w \bar{w}\right) .
\end{align}
as our parametrization for the coordinates and on-shell momentum   \cite{He:2019jjk,He:2020ifr}.
The only saddle point is at $(z,\bar{z}) = (w,\bar{w})$. Making the following change of variables $(y_1,y_2) = \left(\frac{1}{2}(z+ \bar{z}),-\frac{i}{2}(z - \bar{z})\right)$ brings the exponentials into Gaussian form, which integrates to
\begin{align}
\frac{1}{2} \int \mathrm{d}^2 z \, e^{- i \frac{\omega_p r}{2} (z - w)(\bar{z} - \bar{w})} = \int \mathrm{d} y_1 \mathrm{d} y_2 \, e^{- i \frac{\omega_p r}{2} (y_1^2 + y_2^2)} = \frac{(2 \pi) (-i)}{r \omega} .
\end{align}
Substituting into the mode expansion for $\phi(x)$ yields the leading term in the large $r$ limit
\begin{align}
\label{eq:phisaddle}
\phi^{(0)}(u,\hat{\vec{n}}) = \frac{i}{(8 \pi^2)} \int_0^{+\infty} \mathrm{d} \omega \, \left[b^{\dagger}(\omega,z_{\hat{\vec{n}}},\bar{z}_{\hat{\vec{n}}}) e^{i \frac{\omega u}{2}} - b(\omega,z_{\hat{\vec{n}}},\bar{z}_{\hat{\vec{n}}}) e^{-i \frac{\omega u}{2}} \right].
\end{align}
When derivatives in $z$ (or $\bar{z}$) acts on field operators expressions, one can write\footnote{Technically, we take first the saddle point estimate of the quantum field so that it will localize on its point particle expression.}
\begin{align}
\label{eq:dphisaddle}
\partial_z \phi^{(0)}(u,\hat{\vec{n}}) &= \frac{i}{(8 \pi^2)} \int_0^{+\infty} \mathrm{d} \omega_p \, \left[\partial_{z_{\hat{\mathbf{n}}}} b^{\dagger}(\omega_p,z_{\hat{\vec{n}}},\bar{z}_{\hat{\vec{n}}}) e^{i \frac{\omega_p u}{2}} - \partial_{z_{\hat{\mathbf{n}}}} b(\omega_p,z_{\hat{\vec{n}}},\bar{z}_{\hat{\vec{n}}}) e^{-i \frac{\omega_p u}{2}}\right] 
\end{align}
for the saddle point estimate of $\partial_z \phi^{(0)}$ where the derivative acts on the creation and annihilation operators localized around $(w, \bar{w}) = (z_{\hat{\vec{n}}},\bar{z}_{\hat{\vec{n}}})$. 

Inserting the saddle point estimate of $\phi(x)$ into the definition for the scalar ANEC operator at infinity, we find
\begin{align}
\mathcal{E}_{\text{scalar}}(\hat{\vec{n}}) 
&= \int \widetilde{\mathrm{d}^{3} p} \; \omega_{p} \, :b^{\dagger}(\vec{p}) b(\vec{p}): \delta^2(z_{\hat{\vec{n}}} - z_{\hat{\vec{p}}}) ,
\end{align}
where we have set to zero all contributions proportional to $\delta(\omega_{p_1} + \omega_{p_2})$ since the only cases that these constraints can be satisfied correspond to $\omega_{p_1} \to 0^{\pm}$ and $\omega_{p_2} \to 0^{\mp}$, which are unphysical. The action of $\mathcal{E}_{\text{scalar}}(\hat{\vec{n}})$ on an on-shell state $\ket{\vec{X}} = \ket{\vec{p}_1 ... \vec{p}_n}$ gives
\begin{align}
\mathcal{E}_{\text{scalar}}(\hat{\vec{n}}) \ket{\vec{X}} = \sum_{i = 1}^n \left(\omega_i\right) \delta^2(z_{\hat{\vec{n}}} - z_{\hat{\vec{p}}_i}) \ket{\vec{X}} ,
\end{align}
which is natural for an observer located at spatial infinity who is measuring the energy flux along the retarded time $u$ in flat null coordinates. Note that on-shell scattering states are eigenstates of the ANEC operator in the detector limit whose eigenvalues are weight functions similar to equation \eqref{eq:ANECWeight}.

\begin{figure}[h]
\centering
\begin{subfigure}{.45\textwidth}
  \centering
  \includegraphics[width=0.75\textwidth]{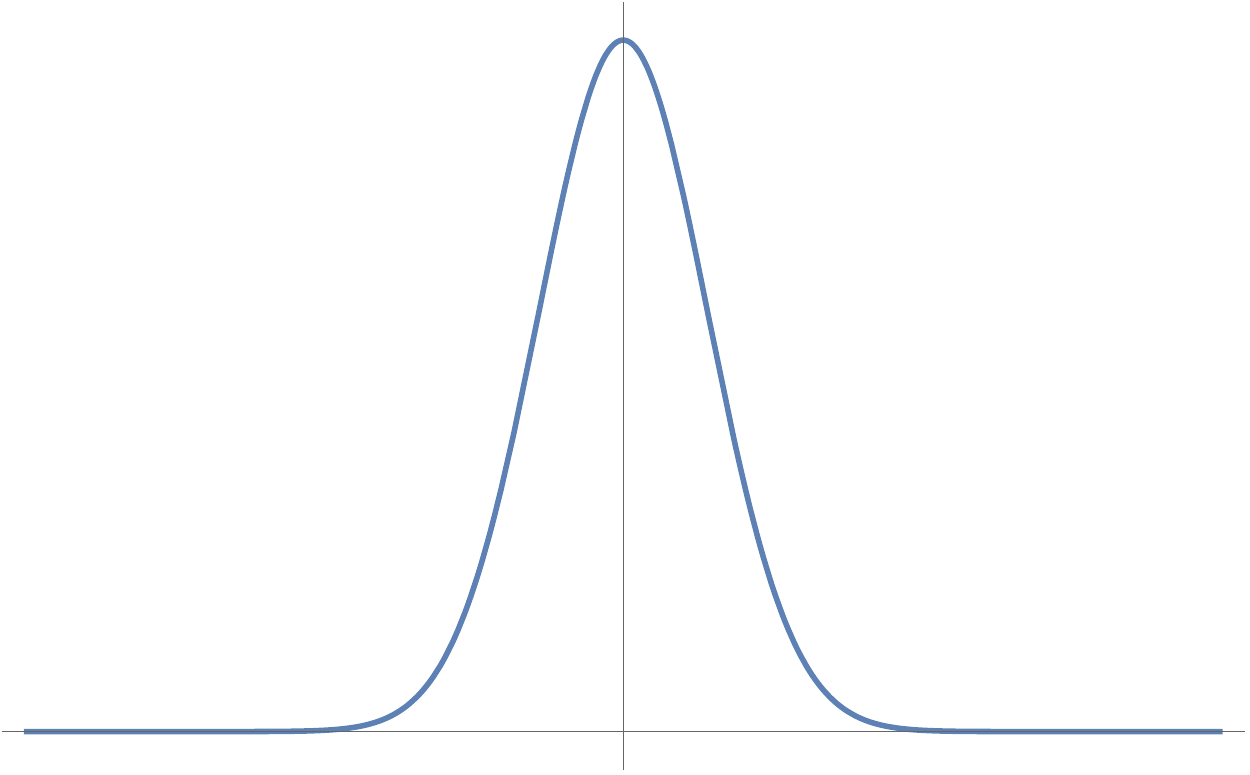}
  \caption{Standard Gaussian shape peaked around $\omega_{\tilde{q}} = \omega_{\tilde{p}}$, which converges to $\delta(\omega_{\tilde{q}} - \omega_{\tilde{p}})$ in the limit of zero width.}
  \label{fig:1a}
\end{subfigure}%
\quad
\begin{subfigure}{.45\textwidth}
  \centering
  \includegraphics[width=0.85\textwidth]{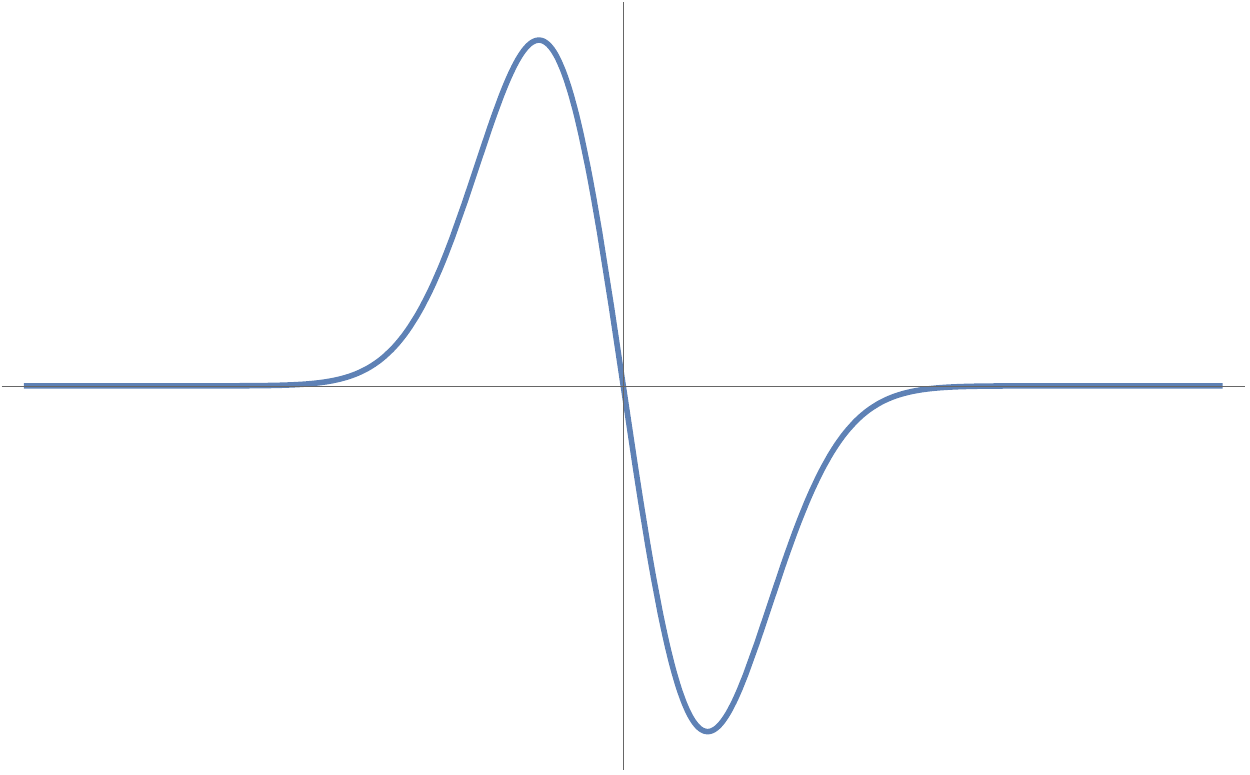}
  \caption{Dipole profile centered around $\omega_{\tilde{q}} = \omega_{\tilde{p}}$, which converges to $\delta^{(1)}(\omega_{\tilde{q}} - \omega_{\tilde{p}})$ in the limit of zero width.}
  \label{fig:1b}
\end{subfigure}
\caption{Comparison of the expectation values of $\bra{\psi_{\vec{q},\text{out}}}  \mathcal{E}_{\text{scalar}}(\hat{\vec{n}}) \ket{\psi_{\vec{p},\text{out}}}$ and $\bra{\psi_{\vec{q},\text{out}}} \mathcal{K}_{\text{scalar}}(\hat{\vec{n}}) \ket{\psi_{\vec{p},\text{out}}}$ as a function of the energy $\omega$.}
\label{fig:1}
\end{figure}

Inserting \eqref{eq:phisaddle} into the definition of the boost energy density flux $\mathcal{K}_{\text{scalar}}(\hat{\vec{n}})$ yields
\begin{align}
&\mathcal{K}_{\text{scalar}}(\hat{\vec{n}}) = \int_{-\infty}^{+\infty} \mathrm{d} u \, (u) \lim_{r \to \infty} r^2 \,\prod_{i=1}^2 \left[\int_0^{+\infty} \mathrm{d} \omega_{p_i} \,\frac{\omega_{p_i}}{2 (8 \pi^2) r} \right]  \nonumber \\
& \times  : \Big[b^{\dagger}(\omega_{p_1}\hat{\vec{n}}) b(\omega_{p_2}\hat{\vec{n}}) e^{- i \frac{(\omega_{p_2} - \omega_{p_1})}{2} u} + b(\omega_{p_1}\hat{\vec{n}}) b(\omega_{p_2}\hat{\vec{n}}) e^{- i \frac{(\omega_{p_1} + \omega_{p_2})}{2} u}  + h.c.\Big] : .
\end{align}
It is convenient to change variables
\begin{align}
\omega_- := \frac{\omega_{p_1} - \omega_{p_2}}{2} \qquad \omega_+ := \frac{\omega_{p_1} + \omega_{p_2}}{2} ,
\end{align}
where $\omega_+ \in [\text{max}\{-\omega_-, \omega_-\},+\infty[$ and $\omega_- \in ]-\infty,+\infty[$ is a convenient slicing of the integration region in the new variables.\footnote{The original frequencies are given by $\omega_{p_1} = (\omega_+ + \omega_-)$ and $\omega_{p_2} = (\omega_+ - \omega_-)$ and the integration measure becomes $\mathrm{d} \omega_{p_1} \mathrm{d} \omega_{p_2}=2 \, \mathrm{d} \omega^{-} \mathrm{d} \omega^{+}$.} In the new variables,\footnote{Here we have assumed, as before, that the other contributions dropped because they are unphysical.} 
\begin{align}
\mathcal{K}_{\text{scalar}}(\hat{\vec{n}}) &= \frac{(-i)}{8}  \int_{-\infty}^{+\infty}  \frac{\mathrm{d} \omega_{-}}{(2 \pi)^3} \delta^{(1)}(\omega_{-}) \int_{\text{max}\{-\omega_-, \omega_-\}}^{+\infty} \mathrm{d} \omega_{+} \, ((\omega_+)^2 - (\omega_-)^2)  \nonumber \\
& \hspace{100pt} \times : \Big[b^{\dagger}\left((\omega_+ + \omega_-)\hat{\vec{n}}\right) b\left((\omega_+ - \omega_-)\hat{\vec{n}}\right) - h.c.\Big] : ,
\end{align}
where we have used the following distributional identity (for $n=1$)
\begin{align}
\int \mathrm{d} u \, (u)^n e^{i \omega_- u} &= (2 \pi) (-i)^n \delta^{(n)}(\omega_-) .
\end{align}
The derivative acting on the energy delta function might seem troubling at first sight. To develop more intuition, we study the action of this operator on on-shell states  $\ket{\vec{X}} = \ket{\vec{p}_1 ... \vec{p}_n}$:
\begin{align}
&\mathcal{K}_{\text{scalar}}(\hat{\vec{n}}) \ket{\vec{X}} 
= \frac{(-i)}{4} \int_{-\infty}^{+\infty}  \frac{\mathrm{d} \omega_{-}}{(2 \pi)^3} \delta^{(1)}(\omega_{-}) \int_{\text{max}\{-\omega_-, \omega_-\}}^{+\infty} \mathrm{d} \omega_{+} \, ((\omega_+)^2 - (\omega_-)^2) \nonumber \\
&\qquad \times \sum_{i=1}^n (2 \pi)^3 \frac{4}{\omega_{p_i}} \delta\left((\omega_+ - \omega_-) - \omega_{p_i}\right) \delta^2(z_{\hat{\vec{n}}} - z_{\hat{\vec{p}}_i}) \ket{\vec{p}_1 ... \underbrace{(\omega_+ + \omega_-)\hat{\vec{n}}}_\text{i-th} ... \vec{p}_n} .
\end{align}
Then one can solve the $\omega_+$ integral
\begin{align}
\mathcal{K}_{\text{scalar}}(\hat{\vec{n}}) \ket{\vec{X}} = 2 \sum_{i=1}^n \delta^2(z_{\hat{\vec{n}}} - z_{\hat{\vec{p}}_i}) \int_{-\omega_{p_i}}^{+\infty}  \mathrm{d} \omega (-i \delta^{(1)}(\omega)) \, \left(\omega_{p_i} + \omega \right) \ket{\vec{p}_1 ... \underbrace{(\omega_{p_i} + \omega)\hat{\vec{n}}}_\text{i-th} ... \vec{p}_n} ,
\label{eqn:K1_action} 
\end{align}
where we have relabelled $\omega_-$ as $\omega$. It is also enlightening to compare the simplest matrix elements of $\mathcal{E}_{\text{scalar}}(\hat{\vec{n}})$ and $\mathcal{K}_{\text{scalar}}(\hat{\vec{n}})$
\begin{align}
\bra{\vec{q}} \mathcal{E}_{\text{scalar}}(\hat{\vec{n}}) \ket{\vec{p}} &= 4 (2 \pi)^3 \delta^2(z_{\hat{\vec{n}}} - z_{\hat{\vec{p}}}) \delta^2(z_{\hat{\vec{q}}} - z_{\hat{\vec{n}}}) \delta(\omega_{q} - \omega_{p}) ,  \nonumber \\
\bra{\vec{q}} \mathcal{K}_{\text{scalar}}(\hat{\vec{n}}) \ket{\vec{p}} &= 2 \delta^2(z_{\hat{\vec{n}}} - z_{\hat{\vec{p}}}) \int_{- \omega_{p}}^{+\infty}  \mathrm{d} \omega (-i \delta^{(1)}(\omega)) \, \left(\omega_{p} + \omega \right) \braket{\vec{q}}{(\omega_{p} + \omega) \hat{\vec{n}}}  \nonumber \\
&= 8 (2 \pi)^3 \delta^2(z_{\hat{\vec{n}}} - z_{\hat{\vec{p}}}) \delta^2(z_{\hat{\vec{q}}} - z_{\hat{\vec{n}}}) (-i \delta^{(1)}(\omega_{q} - \omega_{p})) .
\label{eqn:E_K_exp_values}
\end{align}
It is well-known that in QFT the single contraction $\braket{\vec{q}}{\vec{p}}$ must be interpreted in a distributional sense; in order to properly define such objects one needs to smear them with a well-behaved function (see \cite{Collins:2019ozc,Duncan:1493543} and references therein). In the S-matrix context,  one usually considers outgoing states of the form\footnote{A similar idea applies to the ingoing case.}
\begin{align}
\ket{\psi_{\vec{p},\text{out}}} = \int \widetilde{\mathrm{d}^{3} \tilde{p}} \; \phi_{\vec{p}}(\tilde{\vec{p}}) \ket{\tilde{\vec{p}}} ,
\end{align}
where $\phi_{\vec{p}}(\tilde{\vec{p}})$ is a suitable real momentum wavefunction localized around $\vec{p}$. Note that $\ket{\psi_{\vec{p},\text{out}}} \to \ket{\vec{p}}$ when $\phi_{\vec{p}}(\tilde{\vec{p}}) \to (2 \pi)^3 (2 E_{p}) \delta^3(\vec{p} - \tilde{\vec{p}})$. More generally, when $\phi_{\vec{p}}(\tilde{\vec{p}})$ is sufficiently smooth \eqref{eqn:E_K_exp_values} is well defined 
\begin{align}
\bra{\psi_{\vec{q},\text{out}}}  \mathcal{E}_{\text{scalar}}(\hat{\vec{n}}) \ket{\psi_{\vec{p},\text{out}}} &= 4 (2 \pi)^3 \int\widetilde{\mathrm{d}^{3} \tilde{p}} \;  \int \widetilde{\mathrm{d}^{3} \tilde{q}} \; \phi_{\vec{p}}(\tilde{\vec{p}}) \phi_{\vec{q}}(\tilde{\vec{q}})  \nonumber \\
& \hspace{60pt} \times \delta(\omega_{\tilde{q}} - \omega_{\tilde{p}}) \delta^2(z_{\hat{\vec{n}}} - z_{\hat{\tilde{\vec{p}}}}) \delta^2(z_{\hat{\tilde{\vec{q}}}} - z_{\hat{\vec{n}}})  , \nonumber \\
\bra{\psi_{\vec{q},\text{out}}} \mathcal{K}_{\text{scalar}}(\hat{\vec{n}}) \ket{\psi_{\vec{p},\text{out}}} &= 8 (2 \pi)^3 \int\widetilde{\mathrm{d}^{3} \tilde{p}} \;  \int \widetilde{\mathrm{d}^{3} \tilde{q}} \; \phi_{\vec{p}}(\tilde{\vec{p}}) \phi_{\vec{q}}(\tilde{\vec{q}})  \nonumber \\
& \hspace{60pt} \times (-i \delta^{(1)}(\omega_{\tilde{q}} - \omega_{\tilde{p}})) \delta^2(z_{\hat{\vec{n}}} - z_{\hat{\tilde{\vec{p}}}}) \delta^2(z_{\hat{\tilde{\vec{q}}}} - z_{\hat{\vec{n}}}) .
\label{eqn:E_K_exp_values_reg}
\end{align}
Looking at \eqref{eqn:E_K_exp_values_reg}, the interpretation of $\mathcal{K}_{\text{scalar}}(\hat{\vec{n}})$ is also much more clear: while $\delta(\omega_{\tilde{q}} - \omega_{\tilde{p}})$ can be considered the limit of a suitable localized function around $\omega_{\tilde{q}} = \omega_{\tilde{p}}$, it turns out that $\delta^{(1)}(\omega_{\tilde{q}} - \omega_{\tilde{p}})$ is probing a dipole-like pattern around $\omega_{\tilde{q}} = \omega_{\tilde{p}}$ (see \ref{fig:1a} and \ref{fig:1b} respectively).  

The final operator we consider is $\mathcal{N}_{z,\text{scalar}}(\hat{\vec{n}})$, which after using \eqref{eq:phisaddle} and \eqref{eq:dphisaddle} is
\begin{align}
\mathcal{N}_{z,\text{scalar}}(\hat{\vec{n}}) &= \frac{i}{4(2 \pi)^{3}} \int \mathrm{d} \omega_{p} \, \omega_{p} :\Big[ b^{\dagger}\left(\omega_{p}, z_{\hat{\mathbf{n}}}, \bar{z}_{\hat{\mathbf{n}}}\right) \partial_{z_{\hat{\mathbf{n}}}} b\left(\omega_{p}, z_{\hat{\mathbf{n}}}, \bar{z}_{\hat{\mathbf{n}}}\right) 
\nonumber \\& \hspace{120pt} -  b\left(\omega_{p}, z_{\hat{\mathbf{n}}}, \bar{z}_{\hat{\mathbf{n}}}\right) \partial_{z_{\hat{\mathbf{n}}}} b^{\dagger}\left(\omega_{p}, z_{\hat{\mathbf{n}}}, \bar{z}_{\hat{\mathbf{n}}}\right) \Big]:.
\end{align}
More compactly,
\begin{align}
\mathcal{N}_{z,\text{scalar}}(\hat{\mathbf{n}})=i \int \widetilde{\mathrm{d}^{3} p} \, \delta^{2}\left(z_{\hat{\mathbf{n}}}-z_{\hat{\mathbf{p}}}\right) : \left[ b^{\dagger}(\omega_{p} \hat{\mathbf{n}}) \overset\leftrightarrow{\partial}_{z_{\hat{\mathbf{n}}}} b(\omega_{p} \hat{\mathbf{n}})\right] :  ,
\label{eqn:Nz_scalar}
\end{align}
where
\begin{align}
\frac{i}{2} \partial_{z} = L_{u z}=\frac{i}{r}\left(\partial_{u} x^{\mu}\right)\left(\partial_{z} x^{\nu}\right)\left[p_{\mu} \frac{\partial}{\partial p^{\nu}}-p_{\nu} \frac{\partial}{\partial p^{\mu}}\right]
\end{align}
is the standard orbital angular momentum operator. A similar analysis can be repeated for the antiholomorphic component by flipping $z \leftrightarrow \bar{z}$.

\section{The Spin-1 light-ray operators}

Having found explicit expressions for the light-ray operators for massless scalar theories, we now derive similar results for the light-ray operators for massless spin-1 theories. Using the Hilbert stress tensor $T^{\text{matter}}_{\mu \nu} = - \frac{2}{\sqrt{|g|}} \frac{\delta S^{\text{matter}}}{\delta g^{\mu \nu}}$ for pure gauge theories, one obtains
\begin{align}
T^{\mu \nu}_{\text{photon}} =  F^{\mu \alpha} F^{\,\,\nu}_{\alpha} + \frac{1}{4} \eta^{\mu \nu} F^{\alpha \beta} F_{\alpha \beta}
\end{align}
for Maxwell $U(1)$ theory where $F_{\mu \nu} := \partial_{\mu} A_{\nu} - \partial_{\nu} A_{\mu}$ is the $U(1)$ field strength. The generalization to the non-abelian gauge group $SU(N)$ is straightforward
\begin{align}
T^{\mu \nu}_{\text{gluon}} = 2 \text{Tr}(T^a T^b) \left[(F^{a})^{\mu \alpha} (F^b)^{\,\,\nu}_{\alpha} + \frac{1}{4} \eta^{\mu \nu} (F^{a})^{\alpha \beta} (F^{b})_{\alpha \beta} \right] ,
\label{eqn:YM_stress_tensor}
\end{align}
where $F^a_{\mu \nu} := \partial_{\mu} A^a_{\nu} - \partial_{\nu} A^a_{\mu} + g f^{a b c} A^b_{\mu} A^c_{\nu}$ is the $SU(N)$ field strength. Here, the $SU(N)$ generators are chosen to be hermitian $[T^a,T^b] = i f^{a b c} T^c$ and are normalized according to $\text{Tr}(T^a T^b) = \frac{1}{2} \delta^{a b}$.

To discuss the large $r$ limit of the stress-tensor, we need the large $r$ limit of the gauge field, which requires making a gauge choice, as well as, specifying boundary conditions. We will adopt the radiation gauge  \cite{Strominger:2013lka}
\begin{align}
A_u = 0, \qquad \left.r^{2} \nabla^{\mu} A_{\mu}^{a}(x)\right|_{\mathcal{I}-}=\left.2\left[r^{2} \partial_{u} A_{r}^{a}-\partial_{z} A_{\bar{z}}^{a}-\partial_{\bar{z}} A_{z}^{a}\right]\right|_{\mathcal{I}^{-}}=0
\label{eqn:radiation_gauge}
\end{align}
and we require the gauge field components to satisfy the following falloff conditions at infinity
\begin{align}
A_z (u,r,z,\bar{z}) = A_z^{(0)} (u,z,\bar{z}) + \mathcal{O}\left(\frac{1}{r}\right) , \qquad A_r(u,r,z,\bar{z}) = \frac{1}{r^2} A_r^{(2)}(u,z,\bar{z}) + \mathcal{O}\left(\frac{1}{r^3}\right) ,
\end{align}
which are needed to have a finite energy and angular momentum flux at $\mathcal{I}^+$. The stress tensor components relevant for our family of light-ray operators \eqref{eqn:YM_stress_tensor} are
\begin{align}
T_{u u} &=  \frac{4}{r^2} (\partial_u  A_z^{(0),a}) (\partial_u  A_{\bar{z}}^{(0),a}) ,  \\
T_{u z} &= \frac{2}{r^2} (\partial_u  A_{z}^{(0),a}) (2 \partial_{[z}  A_{\bar{z}]}^{(0),a} + g f^{a b c} A_{z}^{(0),b} A_{\bar{z}}^{(0),c} ) + \frac{2}{r^2} (\partial_u  A_{z}^{(0),a}) (\partial_{u}  A_{r}^{(2),a}) , \\
T_{u \bar{z}} &=  \frac{2}{r^2}  (\partial_u  A_{\bar{z}}^{(0),a}) (2 \partial_{[\bar{z}}  A_{z]}^{(0),a} + g f^{a b c} A_{\bar{z}}^{(0),b} A_{z}^{(0),c} ) + \frac{2}{r^2} (\partial_u  A_{\bar{z}}^{(0),a}) (\partial_{u}  A_{r}^{(2),a}) 
\label{eqn:T_gluon}
\end{align}
up to terms of order $\mathcal{O}(1/r^3)$.

We assume that we can work perturbatively with energies above the confinement scale $\Lambda > 0$  and can formally talk about asymptotic gluon states. We use the following parametrization and projection of the polarization vectors
\begin{align}
\epsilon^{+, \mu}(\vec{q}) &= \frac{1}{\sqrt{2}} (\bar{z}_{q},1,-i,-\bar{z}_{q}), \qquad \epsilon^{-, \mu}(\vec{q}) = \frac{1}{\sqrt{2}} (z_{q},1,i,-z_{q}), \nonumber \\
(\partial_z x^{\mu}) \epsilon_{\mu}^{+}(\vec{q}) &= 0 = (\partial_{\bar{z}} x^{\mu}) \epsilon_{\mu}^{-}(\vec{q}), \qquad (\partial_z x^{\mu}) \epsilon_{\mu}^{-}(\vec{q}) = -\frac{r}{\sqrt{2}} = (\partial_{\bar{z}} x^{\mu}) \epsilon_{\mu}^{+}(\vec{q})
\end{align}
and the canonical commutation relations
\begin{align}
[a_{\sigma}(\vec{p}),a^{\dagger}_{\sigma^{\prime}}(\vec{q})] = \frac{4(2\pi)^3}{\omega_p} \delta_{\sigma \sigma^{\prime}}\delta(\omega_p-\omega_q) \delta^2(z_{\hat{\vec{p}}}-z_{\hat{\vec{q}}}) .
\end{align}
Starting with the gluon-ANEC operator at null infinity, we find
\begin{align}
&\mathcal{E}_{\text{gluon}}(\hat{\vec{n}}) = \int \widetilde{\mathrm{d}^{3} p} \,\, \omega_{p} \, \delta^2(z_{\hat{\vec{n}}} - z_{\hat{\vec{p}}}) \sum_{\sigma = \pm 1}: \Big[a^{\dagger,a}_{\sigma}(\omega_{p} \hat{\vec{n}}) a^a_{\sigma}(\omega_{p} \hat{\vec{n}}) \Big] : ,
\end{align}
where the saddle point estimates for $A_z^{(0),a}$ and $A_{\bar{z}}^{(0),a}$ are
\begin{align}
A_z^{(0),a}(u,\hat{\vec{n}}) &= \frac{-i}{(8 \sqrt{2} \pi^2)} \int_0^{+\infty} \mathrm{d} \omega_p   \, \left[a^{\dagger,a}_-(\omega_p \hat{\vec{n}}) e^{i \frac{\omega_p u}{2}} - a^{a}_+(\omega_p \hat{\vec{n}}) e^{-i \frac{\omega_p u}{2}} \right],  \\
A_{\bar{z}}^{(0),a}(u,\hat{\vec{n}}) &= \frac{-i}{(8 \sqrt{2} \pi^2)} \int_0^{+\infty} \mathrm{d} \omega_p \, \left[a^{\dagger,a}_+(\omega_p \hat{\vec{n}}) e^{i \frac{\omega_p u}{2}} - a^{a}_-(\omega_p \hat{\vec{n}}) e^{-i \frac{\omega_p u}{2}} \right] .
\end{align}
Analogous to the scalar case, on-shell gluon states $\ket{\vec{X}} = \ket{\{\vec{p}_1, \sigma_1\},...,\{\vec{p}_n, \sigma_n\}}$ are eigenstates of the gluon-ANEC at infinity
\begin{align}
&\mathcal{E}_{\text{gluon}}(\hat{\vec{n}}) \ket{\vec{X}} = \sum_{i = 1}^n \left(\omega_i\right) \delta^2(z_{\hat{\vec{n}}} - z_{\hat{\vec{p}}_i}) \ket{\vec{X}} .
\label{eqn:Egluon}
\end{align}
The gluon boost energy density operator $\mathcal{K}_{\text{gluon}}(\hat{\vec{n}})$ follows straightforwardly 
\begin{align}
\mathcal{K}_{\text{gluon}}(\hat{\vec{n}}) &= \frac{(-i) }{4}  \int_{-\infty}^{+\infty}  \frac{\mathrm{d} \omega_{-}}{(2 \pi)^3} \delta^{(1)}(\omega_{-}) \int_{\text{max}\{-\omega_-, \omega_-\}}^{+\infty} \mathrm{d} \omega_{+} \, ((\omega_+)^2 - (\omega_-)^2)  \nonumber \\
& \hspace{90pt} \times \sum_{\sigma = \pm 1}: \Big[a_{\sigma}^{a,\dagger}\left((\omega_+ + \omega_-)\hat{\vec{n}}\right) a^a_{\sigma}\left((\omega_+ - \omega_-)\hat{\vec{n}}\right)\Big] : .
\end{align}
These operators are especially simple; the only fundamental difference between \\ $\{\mathcal{E}_\text{gluon}, \mathcal{K}_\text{gluon}\}$ and $\{\mathcal{E}_\text{scalar}, \mathcal{K}_\text{scalar}\}$ is the sum over helicities.

On the other hand, the spin-1 angular momentum flux operators are complicated by boundary terms, which come from integrating by parts in $u$. Explicitly,
\begin{align}
&\int \mathrm{d} u \lim_{r \to +\infty} r^2 T_{u z} \nonumber \\
&\qquad=\int \mathrm{d} u \Big\{4 (\partial_u  A_{z}^{(0),a}) ( \partial_{z}  A_{\bar{z}}^{(0),a}) \Big\} - 2 A_{z}^{(0),a} (-\partial_{u} A_{r}^{(2),a} + \partial_{z} A_{\bar{z}}^{(0),a} + \partial_{\bar{z}} A_{z}^{(0),a}) \Big|_{\mathcal{I}^+_+} , \nonumber \\
&\int \mathrm{d} u  \lim_{r \to +\infty} r^2 T_{u \bar{z}} \nonumber \\ 
&\qquad=\int \mathrm{d} u \Big\{4 (\partial_u  A_{\bar{z}}^{(0),a}) ( \partial_{\bar{z}}  A_{z}^{(0),a})\Big\} - 2 A_{\bar{z}}^{(0),a} (-\partial_{u} A_{r}^{(2),a} + \partial_{z} A_{\bar{z}}^{(0),a} + \partial_{\bar{z}} A_{z}^{(0),a}) \Big|_{\mathcal{I}^+_+} ,
\label{eqn:T_uz_final}
\end{align}
where we have used the leading $u$-equation of motion
\begin{align}
r^2 & \mathcal{D}_{\mu} F^{\mu}_{\,\,\,u} \nonumber \\
&= 2 \partial_u \left[- \partial_u A_r^{(2),a} + \partial_z A^{(0),a}_{\bar{z}} + \partial_{\bar{z}} A^{(0),a}_z \right] + 2 g f^{a b c} \left[ A^{(0),b}_z \partial_u A^{(0),c}_{\bar{z}} + A^{(0),b}_{\bar{z}} \partial_u A^{(0),c}_{z} \right]  = 0 .
\label{eqn:u_eom}
\end{align}
It is worth remarking that the non-abelian contribution appears through the boundary term \eqref{eqn:T_uz_final}; this is similar to the case when the light-sheet is placed in the bulk (see appendix C). For the first contribution to the angular momentum flux density, we use the identity
\begin{align}
&4 (\partial_u  A_{z}^{(0),a}) ( \partial_{z}  A_{\bar{z}}^{(0),a}) 
\nonumber \\ & \qquad\qquad  =2 (\partial_u  A_{z}^{(0),a}) ( \partial_{z}  A_{\bar{z}}^{(0),a}) -2 (\partial_{z} \partial_u  A_{z}^{(0),a}) A_{\bar{z}}^{(0),a} 
+ 2 \partial_{z} \left[ (\partial_u  A_{z}^{(0),a}) A_{\bar{z}}^{(0),a} \right] ,
\end{align}
just to isolate the contribution coming from the total derivative. Regarding the saddle point estimates for $\partial_z A_z^{(0),a}$ and $\partial_z A_{\bar{z}}^{(0),a}$, we have
\begin{align}
&\partial_z A_z^{(0),a}(u,\hat{\vec{n}}) = \frac{-1}{(4 \sqrt{2} \pi^2)} \int_0^{+\infty} \mathrm{d} \omega_p \, \Big\{ e^{i \frac{\omega_p u}{2}} L_{u z} a^{\dagger,a}_-(\omega_p,z_{\hat{\vec{n}}},\bar{z}_{\hat{\vec{n}}}) -e^{-i \frac{\omega_p u}{2}} L_{u z} a^{a}_+(\omega_p,z_{\hat{\vec{n}}},\bar{z}_{\hat{\vec{n}}}) \Big\} , \nonumber \\
&\partial_z A_{\bar{z}}^{(0),a}(u,\hat{\vec{n}}) = \frac{-1}{(4 \sqrt{2} \pi^2)} \int_0^{+\infty} \mathrm{d} \omega_p \, \Big\{ e^{i \frac{\omega_p u}{2}} L_{u z} a^{\dagger,a}_+(\omega_p,z_{\hat{\vec{n}}},\bar{z}_{\hat{\vec{n}}}) -e^{-i \frac{\omega_p u}{2}} L_{u z} a^{a}_-(\omega_p,z_{\hat{\vec{n}}},\bar{z}_{\hat{\vec{n}}}) \Big\} .
\end{align}

This gives the following representation for the angular momentum flux density
\begin{itemize}
\item An orbital angular momentum contribution 
\begin{align}
\mathcal{N}^{\text{orb}}_{z,\text{gluon}}(\hat{\vec{n}}) &= i \int \widetilde{\mathrm{d}^{3} p} \; \delta^2(z_{\hat{\vec{n}}} - z_{\hat{\vec{p}}}) \, \sum_{\sigma = \pm 1}:  \Big[a^{\dagger,a}_{\sigma}(\omega_{p} \hat{\vec{n}}) \overset\leftrightarrow{\partial}_{z_{\hat{\mathbf{n}}}} a^a_{\sigma}(\omega_{p} \hat{\vec{n}}) \Big] : .
\end{align}
\item A spin type contribution
\begin{align}
\mathcal{N}^{\text{spin}}_{z,\text{gluon}}(\hat{\vec{n}}) &=  - i \int \widetilde{\mathrm{d}^{3} p} \; \delta^2(z_{\hat{\vec{n}}} - z_{\hat{\vec{p}}})  \, \sum_{\sigma = \pm 1} \sigma : \partial_{z_{\hat{\vec{n}}}} \left[ a^{\dagger,a}_{\sigma}(\omega_{p} \hat{\vec{n}}) a^a_{\sigma}(\omega_{p} \hat{\vec{n}})\right] : \nonumber \\
&=  i \int \widetilde{\mathrm{d}^{3} p} \; \partial_{z_{\hat{\vec{n}}}}\delta^2(z_{\hat{\vec{n}}} - z_{\hat{\vec{p}}})  \, \sum_{\sigma = \pm 1} \sigma : \left[ a^{\dagger,a}_{\sigma}(\omega_{p} \hat{\vec{n}}) a^a_{\sigma}(\omega_{p} \hat{\vec{n}})\right] : .
\end{align}
This piece is expected from the spin structure of the point particle stress tensor  \cite{Bailey:1975fe}. Indeed the covariant spin matrix reads  \cite{Choi:2019rlz} (with $\sigma$ helicity)
\begin{align}
S_{\mu \nu} = \frac{\sigma}{p^0} \left(
\begin{array}{cccc}
 0 & 0 & 0 & 0 \\
 0 & 0 & p^3 & -p^2 \\
 0 & -p^3 & 0 & p^1 \\
 0 & p^2 & -p^1 & 0 \\
\end{array}
\right) = \left(
\begin{array}{cccc}
 0 & 0 & 0 & 0 \\
 0 & 0 & \frac{\sigma (1-z \bar{z})}{1 + z \bar{z}} & \frac{i \sigma (z-\bar{z})}{1 + z \bar{z}} \\
 0 & -\frac{\sigma (1-z \bar{z})}{1 + z \bar{z}} & 0 & \frac{\sigma (z+\bar{z})}{1 + z \bar{z}} \\
 0 & -\frac{i \sigma (z-\bar{z})}{1 + z \bar{z}} & -\frac{\sigma (z+\bar{z})}{1 + z \bar{z}} & 0 \\
\end{array}
\right) 
\end{align}
and the component $S_{z \bar{z}} = \frac{1}{r^2}(\partial_z x^{\mu})(\partial_{\bar{z}} x^{\nu}) S_{\mu \nu} = i \frac{\sigma}{2}$ enters in the total hard angular momentum operator as shown in appendix A of   \cite{Pasterski:2015tva}.
\end{itemize}
Please note that the splitting of angular momentum into an orbital and spin part is not gauge invariant, and it is done here only for convenience in analyzing different terms.
The boundary terms in \eqref{eqn:T_uz_final} are associated with soft contributions\footnote{Similar boundary terms were also observed in   \cite{He:2020ifr}, although He and Mitra considered the total angular momentum flux charge.}. Using the equations of motion, we can write 
\begin{align}
-2(-\partial_{u} A_{r}^{(2)} + \partial_{z} A_{\bar{z}}^{(0)} + \partial_{\bar{z}} A_{z}^{(0)}) \Big|_{\mathcal{I}^+_+} = 2 \int_{-\infty}^{+\infty} \mathrm{d} u \, [A_z^{(0)}, \overset\leftrightarrow{\partial}_{u} A_{\bar{z}}^{(0)}] .
\label{eqn:boundary}
\end{align}
This makes it clear that for pure Maxwell theory (which is conformal) this term vanishes identically. Using the saddle point we obtain
\begin{align}
-2 (-\partial_{u} A_{r}^{(2),a} + & \partial_{z} A_{\bar{z}}^{(0),a} + \partial_{\bar{z}} A_{z}^{(0),a}) \Big|_{\mathcal{I}^+_+} \nonumber\\ &=  - 2 i g f^{a b c} \int \widetilde{\mathrm{d}^3 p} \; \sum_{\sigma = \pm 1} \left[a^{\dagger,b}_{\sigma}(\vec{p}) a^{c}_{\sigma}(\vec{p})\right] \delta^2(z_{\hat{\vec{n}}} - z_{\hat{\vec{p}}}) ,
\label{eqn:boundary2}
\end{align}
where in the last line we recognize the gluon number density operator
\begin{align}
\rho^a_{\text{gluon}}(\vec{p}) := -i f^{a b c} \sum_{\sigma = \pm 1} \left[a^{\dagger,b}_{\sigma}(\vec{p}) a^{c}_{\sigma}(\vec{p})\right] ,
\label{eqn:gluon_number}
\end{align}
which contributes to the hard part of the non-abelian (large gauge) charge  \cite{Gonzo:2019fai,He:2020ifr}. Defining the boundary value of the field  \cite{Strominger:2013lka,He:2020ifr}
\begin{align}
C^a_z(z,\bar{z}) := A_{z}^{a}(u=+\infty,z,\bar{z}) \qquad C^a_{\bar{z}}(z,\bar{z}) := A_{\bar{z}}^{a}(u=+\infty,z,\bar{z})
\end{align}
the total boundary contribution to the angular momentum flux becomes
\begin{align}
&- 2 A_z^{(0),a} (-\partial_{u} A_{r}^{(2),a} + \partial_{z} A_{\bar{z}}^{(0),a} + \partial_{\bar{z}} A_{z}^{(0),a}) \Big|_{\mathcal{I}^+_+}  
\nonumber\\ & \hspace{100pt} = 2 g  C^a_z(z,\bar{z}) \int \widetilde{\mathrm{d}^3 p} \, \rho^a_{\text{gluon}}(\vec{p}) \delta^2(z_{\hat{\vec{n}}} - z_{\hat{\vec{p}}}) .
\end{align}
This contribution is soft, since it contains the soft mode $C^a_z(z,\bar{z})$\footnote{Please notice that as in scalar case, we could have changed the boundary conditions to set $C^a_z(z,\bar{z})=0$ in order to allow only hard modes.}. It is also contains the hard gluon charge, which might seem strange at first. We can gain some insight by turning on a matter source $j^{\mu}$, which modifies the $u$-equation of motion and therefore the final density of angular momentum flux
\begin{align}
\int \mathrm{d} u \lim_{r \to +\infty} r^2 T_{u z} &= \int \mathrm{d} u \Big\{4 (\partial_u  A_{z}^{(0),a}) ( \partial_{z}  A_{\bar{z}}^{(0),a}) \Big\} + \int \mathrm{d} u \, A_z^{(0),a} j_u^{a,(2)} + C_z^{a} q_H^a ,\nonumber \\
\int \mathrm{d} u  \lim_{r \to +\infty} r^2 T_{u \bar{z}} &= \int \mathrm{d} u \Big\{4 (\partial_u  A_{\bar{z}}^{(0),a}) ( \partial_{\bar{z}}  A_{z}^{(0),a})\Big\} + \int \mathrm{d} u \, A_{\bar{z}}^{(0),a} j_u^{a,(2)} + C_{\bar{z}}^{a} q_H^a ,
\label{eqn:T_uz_matter}
\end{align}
where
\begin{align}
q_H(z,\bar{z}) :=  \int_{-\infty}^{+\infty} \mathrm{d} u \,\left[ 2 [A_z^{(0)}, \overset\leftrightarrow{\partial}_{u} A_{\bar{z}}^{(0)}] + j_u^{(2)}\right] 
\end{align}
is the density of the hard color charge. These terms were first found by Ashtekar et al.  \cite{Ashtekar:2017wgq,Ashtekar:2017ydh} in the abelian spin 1 case (coupled with a matter distribution) and they are related to mixing of the so-called radiative modes (i.e. our $A_{z}^{(0),a}$ and $A_{\bar{z}}^{(0),a}$) and of coulombic modes (i.e. the hard charge distribution, potential modes). A similar expression holds for $\mathcal{N}_{\bar{z},\text{gluon}}(\hat{\vec{n}})$ which can be obtained by exchanging $z \leftrightarrow \bar{z}$ and flipping all helicity terms $(+) \leftrightarrow (-)$.

\section{The Spin-2 light-ray operators}

It is well-known that the equivalence principle prevents a universal definition of a stress tensor for the gravitational field. One can try to define various nontensorial  objects called ``pseudotensors'' which when linearized transform as Lorentz tensor. However, all such definitions are not gauge invariant (a notable example is the Landau-Lifshitz pseudotensor   \cite{Landau:1982dva}). 
In this paper, we will only consider linearized gravity, i.e. we split the metric tensor as $g_{\mu \nu} = \eta_{\mu \nu} + h_{\mu \nu}$ and assume $|h_{\mu \nu}(x)| \ll 1$. In the physically relevant situation of gravitational waves propagating in an asymptotically flat region, a suitable notion of a \emph{gauge-invariant} stress tensor for long-wavelength modes can be defined using the so-called Isaacson averaging procedure  \cite{Isaacson:1967zz,Isaacson:1968zza}. In our conventions, the effective stress tensor for gravitational waves is   \cite{Maggiore:1900zz}
\begin{align}
T_{\mu \nu}^{\text{eff},\text{GW}} = \frac{\kappa^2}{32 \pi G} \langle \partial_{\mu} h^{\alpha \beta} \partial_{\nu} h_{\alpha \beta}\rangle ,
\label{eqn:GW_stress_tensor}
\end{align}
where $\langle \cdot \rangle$ stands for the average over short-wavelength graviton modes.\footnote{We stress that the gauge invariance of \eqref{eqn:GW_stress_tensor} holds only under the average prescription  \cite{Maggiore:1900zz}.} It is possible to extend the Isaacson construction of an effective gravitational wave stress tensor to general classical theories of gravity  \cite{Stein:2010pn}: for example, the leading order contribution at infinity is proportional to the Ricci mode for $f(R)$ theories   \cite{Berry:2011pb}, conversely the spectrum is unchanged for most higher derivative theories of gravity since higher derivative terms usually drop off  \cite{Stein:2010pn}. It is worth stressing that this stress-tensor is defined canonically a la' Noether, and therefore it represents only the physical energy and the momentum flux gravitational wave contribution whereas the angular momentum requires a different analysis. 

Asymptotically flat geometries have the following large $r$ asymptotic expansion near $\mathcal{I}^+$ in the Bondi gauge  \cite{Bondi:1962px,Sachs:1962wk}
\begin{align}
\mathrm{d} s^2 &= \mathrm{d} v^2 + 2 dv dr - 2 r^2 \gamma_{\zeta \bar{\zeta}} \mathrm{d} \zeta \mathrm{d} \bar{\zeta} \nonumber \\
& - \frac{2 m_B}{r} \mathrm{d} v^2 - r C_{\zeta\zeta} \mathrm{d} \zeta^2 - r C_{\bar{\zeta} \bar{\zeta}} \mathrm{d} \bar{\zeta}^2 - D^\zeta C_{\zeta \zeta} \mathrm{d} v \mathrm{d} \zeta - D^{\bar{\zeta}} C_{\bar{\zeta} \bar{\zeta}} \mathrm{d} v \mathrm{d} \bar{\zeta}  \nonumber \\
&- \frac{1}{r} \left[\frac{4}{3} N_{\zeta} - \frac{1}{4} \partial_\zeta (C_{\zeta \zeta} C^{\zeta \zeta}) \right] \mathrm{d} v \mathrm{d} \zeta - \frac{1}{r} \left[\frac{4}{3} N_{\bar{\zeta}} - \frac{1}{4} \partial_{\bar{\zeta}} (C_{\bar{\zeta} \bar{\zeta}} C^{\bar{\zeta} \bar{\zeta}}) \right] \mathrm{d} v \mathrm{d} \bar{\zeta} + ...
\label{eqn:Bondi_gauge_expansion}
\end{align}
where the leading contribution comes from Minkowski spacetime and subleading contributions encode deviations from the flat background. Bondi coordinates are defined as
\begin{align}
x^{\mu} = \left(v + r, r \frac{\zeta + \bar{\zeta}}{(1 + \zeta \bar{\zeta})}, - i r \frac{\zeta - \bar{\zeta}}{(1 + \zeta \bar{\zeta})},r \frac{1 - \zeta \bar{\zeta}}{(1 + \zeta \bar{\zeta})}\right) \,,
\end{align}
where $v = t - r$ is the retarded time. In this formulation $m_B(v,\zeta,\bar{\zeta})$ is the Bondi mass aspect,  $C_{\alpha \beta}(v,\zeta,\bar{\zeta})$ is the shear tensor and $(N_{\zeta}(v,\zeta,\bar{\zeta}),N_{\bar{\zeta}}(v,\zeta,\bar{\zeta}))$ is the Bondi angular momentum aspect. Here we use the conventions of Pasterski-Strominger-Zhiboedev  \cite{Pasterski:2015tva}, which differs from Barnich-Troessaert  \cite{Barnich:2011mi}, Hawking-Perry-Strominger   \cite{Hawking:2016sgy} or Flanagan-Nichols  \cite{Flanagan:2015pxa}. Since different conventions change the definition of the light-ray operators, this is of particular relevance here.\footnote{See also   \cite{Compere:2018ylh} where some of the relations between different conventions are spelled out in detail.}

Spaces that admit expansions of the form \eqref{eqn:Bondi_gauge_expansion} are called Christodoulou-Klainerman (CK) spaces in the literature  \cite{Strominger:2013jfa,Christodoulou:1993uv}. Physically, the shear tensor encodes gravitational radiation analogously to Maxwell field strength\footnote{In light-cone gauge, $A_v = 0$.} $F_{v \zeta} = \partial_{v} A_{\zeta}$. The Bondi news is defined to be   \cite{Strominger:2013jfa}
\begin{align}
N_{\zeta \zeta}(v,\zeta,\bar{\zeta}) = \partial_v C_{\zeta \zeta}(v,\zeta,\bar{\zeta}) .
\end{align}
In these coordinates, 
\begin{align}
C_{\zeta \zeta}(v,\zeta,\bar{\zeta}) = -\kappa \lim_{r \to \infty} \frac{1}{r} h_{\zeta \zeta}(r, v, \zeta,\bar{\zeta}) ,
\end{align}
where $\kappa = \sqrt{32 \pi G}$   \cite{He:2014laa}. The time-time component of the Einstein equations \eqref{eqn:Bondi_gauge_expansion} gives the Bondi mass loss formula
\begin{align}
\partial_v m_B &= \frac{1}{4} \left[D^2_\zeta N^{\zeta \zeta} + D^2_{\bar{\zeta}} N^{\bar{\zeta} \bar{\zeta}} - N_{\zeta \zeta} N^{\zeta \zeta} \right] - 4 \pi G \lim_{r \to \infty} r^2 T^{\text{matter}}_{vv} .
\end{align}
For CK spaces, the asymptotics of the Weyl curvature component $\Psi_2^0$ implies the following conditions  \cite{Strominger:2013jfa}
\begin{align}
m_B(v,\zeta,\bar{\zeta}) \Big|_{\mathcal{I}^{+}_-}=  M^{\text{ADM}}_i(\zeta,\bar{\zeta})&, \qquad m_B(v,\zeta,\bar{\zeta}) \Big|_{\mathcal{I}^{+}_+} = M_f(\zeta,\bar{\zeta}), \nonumber \\
C_{\zeta\zeta} = -2 &D_{\zeta}^2 C(\zeta,\bar{\zeta}) ,
\end{align}
where $C(\zeta,\bar{\zeta})$ labels inequivalent BMS vacua related to each other under action of the supertranslation mode. After integrating over $v$ from $-\infty$ to $+\infty$, we get  \cite{Strominger:2014pwa,He:2014laa}
\begin{align}
\Delta m_B = \frac{1}{2} D_{\zeta}^2 \Delta C^{\zeta \zeta} - 4 \pi G \int_{-\infty}^{+\infty} \mathrm{d} v  \Big\{ \frac{1}{16 \pi G} N_{\zeta \zeta} N^{\zeta \zeta} + \lim_{r \to \infty} r^2 T^{\text{matter}}_{v v} \Big\} ,
\label{eqn:Bondi_eq}
\end{align}
where $\Delta C^{\zeta \zeta}(\zeta,\bar{\zeta}) = C^{\zeta \zeta}(+\infty,\zeta,\bar{\zeta}) - C^{\zeta \zeta}(-\infty,\zeta,\bar{\zeta})$ and $\Delta m_B =  M_f(\zeta,\bar{\zeta}) $ since \\ $M^{\text{ADM}}_i(\zeta,\bar{\zeta}) = 0$ for flat Minkowski spacetime. 
Using supertranslation invariance, one can choose a Bondi frame  \cite{Strominger:2013jfa} where $C^{\zeta \zeta}(-\infty,\zeta,\bar{\zeta}) = 0$ but in general the final vacuum will be non-trivial $C^{\zeta \zeta}(+\infty,\zeta,\bar{\zeta}) \neq 0$. 
Focusing on Bondi news squared term
\begin{align}
\frac{1}{16 \pi G} N_{\zeta \zeta} N^{\zeta \zeta} = \frac{\kappa^2}{16 \pi G} \lim_{r \to \infty} \frac{1}{r^2} (\gamma^{\zeta \bar{\zeta}})^2 (\partial_v h_{\zeta \zeta}(x)) (\partial_v h_{\bar{\zeta} \bar{\zeta}}(x)) ,
\end{align}
we see that it agrees with the (light-cone) time-time component of Isaacson effective gravitational wave stress tensor at infinity $T^{\text{eff},GW}_{v v}$  in these coordinates (albeit this is defined with the implicit averaging over short wavelength modes). The action of \eqref{eqn:Bondi_eq} on an on-shell $n$-graviton state $\ket{\vec{X}} = \ket{\{\vec{p}_1,\sigma_1\} ... \{\vec{p}_n,\sigma_n\}}$ is
\begin{align}
\mathcal{\tilde{E}}_{\text{Bondi}}(\vec{\hat{n}})  \ket{\vec{X}} = \int_{-\infty}^{+\infty} \mathrm{d} v \, \frac{1}{16 \pi G} N_{\zeta \zeta} N^{\zeta \zeta} \ket{\vec{X}} = \sum_{i=1}^n (E_{p_i}) \delta^2(\Omega_{\vec{\hat{n}}} - \Omega_{\vec{\hat{p}}_i}) \ket{\vec{X}} ,
\end{align}
where we used the saddle-point estimate for the free field mode expansion for the linearized graviton mode
\begin{align}
h_{\mu \nu}(x) = \int \widetilde{\mathrm{d}^{3} p} \sum_{\sigma = \pm 2} \left[\epsilon^{\sigma, *}_{\mu \nu}(\vec{p}) a_{\sigma}(\vec{p}) e^{-i p \cdot x} + \epsilon^{\sigma}_{\mu \nu}(\vec{p}) a^{\dagger}_{\sigma}(\vec{p}) e^{i p \cdot x} \right] .
\end{align}
The weight factor of the Bondi news squared term is given by the standard local factor $E_{p}$ in retarded Bondi coordinates, which is expected because here $v=t-r$ is the standard light-cone time.

We can now come back to the equation \eqref{eqn:Bondi_eq} which can be solved in terms of $\Delta C_{\zeta\zeta}$  \cite{Strominger:2014pwa}
\begin{align}
\Delta C_{\zeta \zeta} = \frac{4}{\pi} \int \mathrm{d}^2 \zeta^{\prime} \, \gamma_{\zeta^{\prime} \bar{\zeta}^{\prime}} \frac{\bar{\zeta} - \bar{\zeta}^{\prime}}{\zeta - \zeta^{\prime}} \frac{(1 + \zeta^{\prime} \bar{\zeta})^2}{(1 + \zeta^{\prime} \bar{\zeta}^{\prime}) (1 + \zeta \bar{\zeta})^3} \left(4 \pi G \, \mathcal{E}_{\text{shear-inclusive}}(\zeta^{\prime},\bar{\zeta}^{\prime}) + \Delta M_B(\zeta^{\prime},\bar{\zeta}^{\prime}) \right) ,
\label{eqn:memory}
\end{align}
where we have defined a suitable extension of the ANEC operator to a shear-inclusive ANEC at infinity which includes both matter and gravity contributions
\begin{align}
\mathcal{\tilde{E}}_{\text{shear-inclusive}}(\hat{\vec{n}}) := \int_{-\infty}^{+\infty} \mathrm{d} v  \Big\{\frac{1}{16 \pi G}  (N_{\zeta \zeta}(v,\vec{n})) (\gamma^{\zeta \bar{\zeta}})^2 (N_{\bar{\zeta} \bar{\zeta}}(v,\vec{n})) + \lim_{r \to \infty} r^2 T^{\text{matter}}_{v v}(r,v,\vec{n}) \Big\} ,
\end{align}
whose action on an on-shell $n$-particle state of different massless species $\prod_{\alpha \in \text{species}} \ket{X_{\alpha}}$ is
\begin{align}
\mathcal{\tilde{E}}_{\text{shear-inclusive}}(\hat{\vec{n}}) \prod_{\alpha \in \text{species}} \ket{X_{\alpha}} = \sum_{i = 1}^n E_i \delta^2(\Omega_{\vec{\hat{n}}} - \Omega_{\vec{\hat{p}}_i}) \prod_{\alpha \in \text{species}} \ket{X_{\alpha}} .
\end{align}
As we will prove in the next chapter on gravitational event shapes, expectation values of this operator will be positive-definite for a scattering process where we treat (perturbative) gravity as an EFT, similarly to what happens in the massless spin 0 and spin 1 case. Originally, the shear-inclusive ANEC operator was defined for complete achronal null geodesics (also called null lines) to cure the violations of the averaged null energy condition appearing for linearized graviton perturbations, see the appendix D for more details.

At the quantum level we can promote both sides of \eqref{eqn:memory} to operators and act on an on-shell state composed of massless particles reaching null infinity (i.e. radiation): a direct calculation shows that 
\begin{align}
\Delta C_{\zeta \zeta}  \prod_{\alpha \in \text{species}} \ket{X_{\alpha}}&= \Big[ 16 G \sum_{i = 1}^n \frac{\bar{\zeta} - \bar{\zeta}_i}{\zeta - \zeta_i} \frac{(1 + \zeta_i \bar{\zeta})^2 E_i }{(1 + \zeta_i \bar{\zeta}_i) (1 + \zeta \bar{\zeta})^3}\Big] \prod_{\alpha \in \text{species}} \ket{X_{\alpha}} ,
\end{align}
where $\Delta M_B = 0$ in flat Minkowski picture at $v \to \pm \infty$. This is the leading memory effect due to the radiation flux generated in the scattering process  \cite{Tolish:2014oda}, which can be directly related to  Christodoulou memory effect  \cite{Thorne:1992sdb}. Our results are consistent with the memory being given by the (transverse traceless part of) soft factor  \cite{Strominger:2014pwa,Tolish:2014oda}
\begin{align}
M_{\zeta \zeta} = \frac{1}{r^2}(\partial_{\zeta} x^{\mu}) (\partial_{\zeta} x^{\nu}) \frac{\kappa}{2} \left[\sum_{i = 1}^n \eta_i \frac{(p_{\mu})_i (p_{\nu})_i}{p_i \cdot n}\right]^{\text{TT}} ,
\end{align}
once the normalization factor $1 /(4 \pi r)$ is taken into account.

The angular momentum flux, requires the Einstein equations of the subleading terms in the Bondi gauge expansion  \cite{Pasterski:2015tva}
\begin{align}
\partial_v N_{\zeta} &= \frac{1}{4} \partial_{\zeta}\left[D_{\zeta}^{2} C^{\zeta \zeta}-D_{\bar{\zeta}}^{2} C^{\bar{\zeta} \bar{\zeta}}\right]+\partial_{\zeta} m_{B}  \nonumber \\
&- 8 \pi G \left[\lim_{r \to +\infty} r^2  T_{v \zeta}^{\text{matter}} - \frac{1}{32 \pi G} D_{\zeta} (C_{\zeta \zeta} N^{\zeta \zeta}) - \frac{1}{16 \pi G} C_{\zeta \zeta} D_{\zeta} N^{\zeta \zeta}\right] .
\end{align}
Like the shear-inclusive ANEC at infinity, all angular momentum flux contributions that are quadratic in the fields combine together into a single expression for all massless particles
\begin{align}
\mathcal{\tilde{N}}^{\text{PSZ}}_{\zeta, \text{Bondi}}(\vec{\hat{n}}) :=& \int_{-\infty}^{+\infty} \mathrm{d} v \Big\{ \lim_{r \to +\infty} r^2  T_{v \zeta}^{\text{matter}} + \frac{1}{32 \pi G} (\partial_{\zeta} C_{\zeta \zeta}) (\gamma^{\zeta \bar{\zeta}})^2 (\partial_v C_{\bar{\zeta} \bar{\zeta}}) \nonumber \\
&- \frac{1}{32 \pi G} C_{\zeta \zeta} (\gamma^{\zeta \bar{\zeta}})^2 (\partial_{\zeta} \partial_v C_{\bar{\zeta} \bar{\zeta}}) - \frac{1}{16 \pi G} \partial_{\zeta} (C_{\zeta \zeta} (\gamma^{\zeta \bar{\zeta}})^2  \partial_v C_{\bar{\zeta} \bar{\zeta}}) \Big\} ,
\label{eqn:Nz_PSZ}
\end{align}
where the gravity contributions are expressed only in terms of the radiative data. The saddle point estimate of the first two terms of such contributions above gives the density of orbital angular momentum flux
\begin{align}
\mathcal{\tilde{N}}^{\text{PSZ},\text{orb}}_{\zeta, \text{Bondi}}(\vec{\hat{n}}) &= \frac{1}{32 \pi G} \int_{-\infty}^{+\infty} \mathrm{d} v \Big\{(\partial_{\zeta} C_{\zeta \zeta}) (\gamma^{\zeta \bar{\zeta}})^2 (\partial_v C_{\bar{\zeta} \bar{\zeta}}) - C_{\zeta \zeta} (\gamma^{\zeta \bar{\zeta}})^2 (\partial_{\zeta} \partial_v C_{\bar{\zeta} \bar{\zeta}}) \Big\} \nonumber \\
&= -\frac{1}{2} \int \widetilde{\mathrm{d}^{3} p} \,  \frac{\delta^2(\zeta_{\vec{\hat{n}}}-\zeta_{\vec{\hat{p}}})}{\gamma_{\zeta \bar{\zeta}}} \sum_{\sigma = \pm 2}:  \Big[a^{\dagger}_{\sigma}(E_{p} \hat{\vec{n}}) i  \overset\leftrightarrow{\partial}_{\zeta_{\hat{\mathbf{n}}}} a_{\sigma}(E_{p} \hat{\vec{n}}) \Big] : .
\end{align}
The remaining term gives the spin contribution 
\begin{align}
\mathcal{\tilde{N}}^{\text{PSZ},\text{spin}}_{\zeta, \text{Bondi}}(\vec{\hat{n}}) &=-\frac{i}{2} \int \widetilde{\mathrm{d}^{3} p} \,  \left[ \frac{\delta^2(\zeta_{\vec{\hat{n}}}-\zeta_{\vec{\hat{p}}})}{\gamma_{\zeta \bar{\zeta}}} \right] \sum_{\sigma = \pm 2} \sigma : \partial_{\zeta_{\vec{\hat{n}}}}\Big[a^{\dagger}_{\sigma}(E_{p} \hat{\vec{n}}) a_{\sigma}(E_{p} \hat{\vec{n}})\Big] : \nonumber \\ 
&=\frac{i}{2} \int \widetilde{\mathrm{d}^{3} p} \,   \partial_{\zeta_{\vec{\hat{n}}}}\left[\frac{\delta^2(\zeta_{\vec{\hat{n}}}-\zeta_{\vec{\hat{p}}})}{\gamma_{\zeta \bar{\zeta}}} \right] \sum_{\sigma = \pm 2} \sigma : \Big[a^{\dagger}_{\sigma}(E_{p} \hat{\vec{n}}) a_{\sigma}(E_{p} \hat{\vec{n}})\Big] : .
\end{align}
Other conventions like Hawking-Perry-Strominger (HPS) differ from \eqref{eqn:Nz_PSZ} in total derivative terms in $\zeta$ or $\bar{\zeta}$ of $C_{\zeta \zeta} N^{\zeta \zeta}$ and $C_{\bar{\zeta} \bar{\zeta}} N^{\bar{\zeta} \bar{\zeta}}$, which look like contact/spin terms under a saddle-point analysis similar to the one discuss before. While we are not attempting any kind of rigorous analysis of all these terms in different conventions here, our expectation is that the PSZ convention is the most natural\footnote{Probably also because it makes direct contact with the Weyl tensor component $\Psi_1^0$.} choice which gives the expected spin term of the graviton.  This was explored further in  \cite{Compere:2019gft}, where the most general expression of the total angular momentum flux was analyzed and it was shown that \eqref{eqn:Nz_PSZ} corresponds to the simplest case $\alpha = \beta = 0$ in their notation.

To make contact with the system of coordinates used in the rest of the paper, we re-write the corresponding expressions in flat null coordinates\footnote{These expressions can be derived in other ways. For example, $\mathcal{E}_{\text{GR}}(\hat{\vec{n}})$ and $\mathcal{K}_{\text{GR}}(\hat{\vec{n}})$ can be deduced from the saddle point estimate of the Isaacson effective stress tensor in flat null coordinates.}
\begin{align}
\mathcal{E}_{\text{GR}}(\hat{\vec{n}}) &:= \frac{1}{8 \pi G} \int_{-\infty}^{+\infty} \mathrm{d} u \,(\partial_u C_{zz})(\partial_u C_{\bar{z}\bar{z}}) ,  \quad \mathcal{K}_{\text{GR}}(\hat{\vec{n}}) := \frac{1}{8 \pi G} \int_{-\infty}^{+\infty} \mathrm{d} u \, u (\partial_u C_{zz})(\partial_u C_{\bar{z}\bar{z}}) , \nonumber \\
\mathcal{N}^{\text{PSZ}}_{z, \text{GR}}(\hat{\vec{n}}) &:= \frac{1}{16 \pi G} \int_{-\infty}^{+\infty} \mathrm{d} u \, \left[(\partial_{z} C_{z z}) (\partial_u C_{\bar{z} \bar{z}}) - C_{z z} (\partial_{z} \partial_u C_{\bar{z} \bar{z}})- 2 \partial_{z} (C_{z z} \partial_u C_{\bar{z} \bar{z}})\right] , 
\label{eqn:light-ray_GR}
\end{align}
whose saddle point estimate gives
\begin{align}
\mathcal{E}_{\text{GR}}(\hat{\vec{n}}) &= \int \widetilde{\mathrm{d}^{3} p} \, \, \omega_{p} \,  \delta^2(z_{\hat{\vec{n}}} - z_{\hat{\vec{p}}}) \sum_{\sigma = \pm 2}: \Big[a^{\dagger}_{\sigma}(\omega_{p} \hat{\vec{n}}) a_{\sigma}(\omega_{p} \hat{\vec{n}}) \Big] : , \nonumber \\
\mathcal{K}_{\text{GR}}(\hat{\vec{n}}) &= \frac{(-i)}{4}  \int_{-\infty}^{+\infty}  \frac{\mathrm{d} \omega_{-}}{(2 \pi)^3} \delta^{(1)}(\omega_{-}) \int_{\text{max}\{-\omega_-, \omega_-\}}^{+\infty} \mathrm{d} \omega_{+} \, ((\omega_+)^2 - (\omega_-)^2) \nonumber \\
& \hspace{100pt} \times \sum_{\sigma = \pm 2}: \Big[a_{\sigma}^{\dagger}\left((\omega_+ + \omega_-)\hat{\vec{n}}\right) a_{\sigma}\left((\omega_+ - \omega_-)\hat{\vec{n}}\right) \Big] : , \nonumber \\
\mathcal{N}^{\text{PSZ},\text{orb}}_{z,\text{GR}}(\hat{\vec{n}}) &= i \int \widetilde{\mathrm{d}^{3} p} \, \delta^2(z_{\hat{\vec{n}}} - z_{\hat{\vec{p}}}) \, \sum_{\sigma = \pm 2}:  \Big[a^{\dagger}_{\sigma}(\omega_{p} \hat{\vec{n}}) \overset\leftrightarrow{\partial}_{z_{\hat{\mathbf{n}}}} a_{\sigma}(\omega_{p} \hat{\vec{n}})\Big] : , \nonumber \\
\mathcal{N}^{\text{PSZ},\text{spin}}_{z,\text{GR}}(\hat{\vec{n}}) &= - i \int \widetilde{\mathrm{d}^{3} p} \, \delta^2(z_{\hat{\vec{n}}} - z_{\hat{\vec{p}}}) \,\sum_{\sigma = \pm 2} \sigma :\partial_{z_{\vec{\hat{n}}}} \left[ a^{\dagger}_{\sigma}(\omega_{p} \hat{\vec{n}}) a_{\sigma}(\omega_{p} \hat{\vec{n}}) \right] : \nonumber  \\
&=  i \int \widetilde{\mathrm{d}^{3} p} \, \partial_{z_{\vec{\hat{n}}}}\delta^2(z_{\hat{\vec{n}}} - z_{\hat{\vec{p}}}) \,\sum_{\sigma = \pm 2} \sigma : \left[ a^{\dagger}_{\sigma}(\omega_{p} \hat{\vec{n}}) a_{\sigma}(\omega_{p} \hat{\vec{n}}) \right] : .
\end{align}
Note that the spin operator above differs from the spin-1 counterpart by a factor of 2 since the graviton has helicities $\sigma = \pm 2$.

\section{On-shell light-ray algebra}
\label{sec:onshell_algebra}

In general, computing the commutator algebra of composite (non-local\footnote{The fact that our operators are non-local does not guarantee the convergence of the OPE of such operators in the first place. This problem can be solved in CFT  \cite{Kologlu:2019mfz}, but it is not clear how to generalize it to a general quantum field theory.}) operators is quite subtle unless completely fixed by symmetries. 
For example, with the Schwinger action principle  \cite{Schwinger:1959xd,Schwinger:1962wd,Schwinger:1963zz} we can compute the structure of all commutators of the (covariantly conserved) stress tensor but the commutators of stress tensor components with spatial-spatial indices; such commutators are model-dependent  \cite{Deser:1967zzf}. Since our family of light-ray operators contain either the time-time components or mixed time-spatial components of the stress tensor, it is conceivable that our light-ray operators satisfy a universal algebra   \cite{Huang:2019fog,Huang:2020ycs,Cordova:2018ygx} 
\begin{align}
[L_1(\hat{\vec{n}}_1), L_2(\hat{\vec{n}}_2)] 
= \sum_{L_3} C_{L_1 L_2 L_3}(\hat{\vec{n}}_1,\hat{\vec{n}}_2)\, L_3(\hat{\vec{n}}_2).
\label{eqn:L_algebra}
\end{align}
To make contact with physical detectors at infinity, we study the light ray algebra by taking the difference between the expectation value of Wightman two-point functions:
\begin{align}
\langle [L(\hat{\vec{n}}_1), L^{\prime}(\hat{\vec{n}}_2)] \rangle 
:= \bra{\psi_L} L(\hat{\vec{n}}_1) L^{\prime}(\hat{\vec{n}}_2) \ket{\psi_R} \
- \bra{\psi_L} L^{\prime}(\hat{\vec{n}}_2) L(\hat{\vec{n}}_1) \ket{\psi_R} .
\end{align}
Without loss of generality, we can set the $\vert\psi\rangle$ above to be a single particle state.  In the following calculations we will assume that canonical commutation relations are valid, motivated by the fact we will compute the algebra on a flat section of null infinity thanks to the choice of flat null coordinates. 

\subsection{Spin 0}

The simplest commutator is $\mathcal{P}^{\mu}_{\text{scalar}}(\hat{\vec{n}}_1)$ and $\mathcal{E}_{\text{scalar}}(\hat{\vec{n}}_2)$:
\begin{align}
\bra{\vec{q}} [\mathcal{P}^{\mu}_{\text{scalar}}(\hat{\vec{n}}_1), \mathcal{E}_{\text{scalar}}(\hat{\vec{n}}_2)] \ket{\vec{p}} = 0 .
\end{align}
The vanishing of this commutator -at all order in perturbation theory- is physically interpreted as the statement that measurements of energy and momentum in two localized directions $\hat{\vec{n}}_1$ and $\hat{\vec{n}}_2$ are compatible at the quantum mechanical level. 

The commutator $[\mathcal{K}_{\text{scalar}}(\hat{\vec{n}}_1), \mathcal{E}_{\text{scalar}}(\hat{\vec{n}}_2)]$ is slightly more non-trivial. Using 
\begin{align}
&\bra{\vec{q}} [\mathcal{K}_{\text{scalar}}(\hat{\vec{n}}_1), \mathcal{E}_{\text{scalar}}(\hat{\vec{n}}_2)] \ket{\vec{p}} 
\nonumber\\&\qquad\qquad= \left(\delta^2(z_{\hat{\vec{n}}_2} - z_{\hat{\vec{p}}}) \omega_{p}-\delta^2(z_{\hat{\vec{n}}_2} - z_{\hat{\vec{q}}}) \omega_{q}\right)\bra{\vec{q}} \mathcal{K}_{\text{scalar}}\left(\hat{\vec{n}}_{1}\right)\ket{\vec{p}} ,
\end{align}
one obtains\footnote{The following steps can be made rigorous by smearing the states with appropriate smooth wavefunctions, as explained earlier.}
\begin{align}
\bra{\vec{q}} &[\mathcal{K}_{\text{scalar}}(\hat{\vec{n}}_1), \mathcal{E}_{\text{scalar}}(\hat{\vec{n}}_2)] \ket{\vec{p}}= \nonumber \\
&=8 (2 \pi)^{3} i \underbrace{\left(\omega_{q}-\omega_{p}\right) \delta^{(1)}\left(\omega_{q}-\omega_{p}\right)}_{-\delta\left(\omega_{q}-\omega_{p}\right)} \delta^2(z_{\hat{\vec{n}_1}} - z_{\hat{\vec{n}}_2}) \delta^2(z_{\hat{\vec{n}}_2} - z_{\hat{\vec{p}}}) \delta^2(z_{\hat{\vec{n}}_1} - z_{\hat{\vec{q}}}) \nonumber \\
&=-8(2 \pi)^{3} i \delta\left(\omega_{q}-\omega_{p}\right) \delta^2(z_{\hat{\vec{n}}_1} - z_{\hat{\vec{n}}_2}) \delta^2(z_{\hat{\vec{n}}_2} - z_{\hat{\vec{p}}}) \delta^2(z_{\hat{\vec{n}}_1} - z_{\hat{\vec{q}}}) \nonumber \\
&=-2 i \delta^2(z_{\hat{\vec{n}}_1} - z_{\hat{\vec{n}}_2}) \left[\left(\omega_p  \delta^2(z_{\hat{\vec{n}}_2} - z_{\hat{\vec{p}}})\right) \left( (2 \pi)^{3} 4 \frac{\delta\left(\omega_{q}-\omega_{p}\right)}{\omega_{p}} \delta^2(z_{\hat{\vec{n}}_1} - z_{\hat{\vec{q}}})\right) \right] \nonumber \\
&=-2 i  \delta^2(z_{\hat{\vec{n}}_1} - z_{\hat{\vec{n}}_2}) \bra{\vec{q}}\mathcal{E}_{\text{scalar}}(\hat{\vec{n}}_2) \ket{\vec{p}},
\end{align}
which implies 
\begin{align}
&[\mathcal{K}_{\text{scalar}}(\hat{\vec{n}}_1), \mathcal{E}_{\text{scalar}}(\hat{\vec{n}}_2)] = -2 i  \delta^2(z_{\hat{\vec{n}}_1} - z_{\hat{\vec{n}}_2}) \mathcal{E}_{\text{scalar}}(\hat{\vec{n}}_2) .
\end{align}
Moreover, it is straightforward to check that 
\begin{align}
&[\mathcal{E}_{\text{scalar}}(\hat{\vec{n}}_1), \mathcal{E}_{\text{scalar}}(\hat{\vec{n}}_2)] = 0, \nonumber \\
&[\mathcal{K}_{\text{scalar}}(\hat{\vec{n}}_1), \mathcal{K}_{\text{scalar}}(\hat{\vec{n}}_2)] = 0.
\end{align}
Moving onto the commutators involving the angular momentum flux density operator we need to introduce some smearing in the transverse directions $(z,\bar{z})$ to be able to regularize the expressions, as in \cite{Cordova:2018ygx,Belin:2020lsr}. With this technique we find that 
\begin{align}
&\bra{\vec{q}} [\mathcal{N}_{z,\text{scalar}}(\hat{\vec{n}}_1), \mathcal{E}_{\text{scalar}}(\hat{\vec{n}}_2)] \ket{\vec{p}} \nonumber \\
&= (\omega_p  \delta^2(z_{\hat{\vec{p}}} - z_{\hat{\vec{n}}_2}) - \omega_q  \delta^2(z_{\hat{\vec{q}}} - z_{\hat{\vec{n}}_2})) \bra{\vec{q}} \mathcal{N}_{z,\text{scalar}}(\hat{\vec{n}}_1) \ket{\vec{p}} \nonumber \\
&=-2 i \delta^2(z_{\hat{\vec{n}}_1} - z_{\hat{\vec{n}}_2}) \bra{\vec{q}} \partial_{z}\mathcal{E}_{\text{scalar}}(\hat{\vec{n}}_2)\ket{\vec{p}} + 2 i \partial_{z} \delta^2(z_{\hat{\vec{n}}_1} - z_{\hat{\vec{n}}_2}) \bra{\vec{q}} \mathcal{E}_{\text{scalar}}(\hat{\vec{n}}_2)\ket{\vec{p}} .
\end{align}
Similarly, 
\begin{align}
&\bra{\vec{q}} [\mathcal{N}_{z,\text{scalar}}(\hat{\vec{n}}_1), \mathcal{K}_{\text{scalar}}(\hat{\vec{n}}_2)] \ket{\vec{p}} 
\nonumber \\
&\quad= - 2 i \delta^2(z_{\hat{\vec{n}}_1} - z_{\hat{\vec{n}}_2}) \bra{\vec{q}} \partial_{z}\mathcal{K}_{\text{scalar}}(\hat{\vec{n}}_2)\ket{\vec{p}} 
+ 2 i \partial_{z} \delta^2(z_{\hat{\vec{n}}_1} - z_{\hat{\vec{n}}_2}) \bra{\vec{q}} \mathcal{K}_{\text{scalar}}(\hat{\vec{n}}_2)\ket{\vec{p}} .
\end{align}
The last commutator  $[\mathcal{N}_{z,\text{scalar}}(\hat{\vec{n}}_1), \mathcal{N}_{\bar{z},\text{scalar}}(\hat{\vec{n}}_2)]$ gives
\begin{align}
&\bra{\vec{q}} [\mathcal{N}_{z,\text{scalar}}(\hat{\vec{n}}_1), \mathcal{N}_{\bar{z},\text{scalar}}(\hat{\vec{n}}_2)] \ket{\vec{p}} 
\nonumber \\ 
&\quad = 
	+ 2 i \partial_{z}  
		\delta^2(z_{\hat{\vec{n}}_1} - z_{\hat{\vec{n}}_2}) 
		\bra{\vec{q}} 
			\mathcal{N}_{\bar{z},\text{scalar}}(\hat{\vec{n}}_2)
		\ket{\vec{p}} 
	- 2 i \partial_{\bar{z}}  
		\delta^2( z_{\hat{\vec{n}}_1} - z_{\hat{\vec{n}}_2}) 	
		\bra{\vec{q}} 
			\mathcal{N}_{z,\text{scalar}}(\hat{\vec{n}}_2)
		\ket{\vec{p}} 
	\nonumber \\
	&\qquad 
	-2 i \delta^2(z_{\hat{\vec{n}}_1} - z_{\hat{\vec{n}}_2}) 
	\bra{\vec{q}} 
		\partial_{z} 
			\mathcal{N}_{\bar{z},\text{scalar}}(\hat{\vec{n}}_2) 
	\ket{\vec{p}} .
\end{align}
This concludes the calculation of the scalar light-ray algebra. It is worth stressing that in this case our boundary conditions at infinity naturally select only the hard modes.

\subsection{Spin 1}

The spin 1 abelian case (i.e. Maxwell theory) does not differ much from the spin 0 case except for the presence of the helicity term. The additional helicity term commutes with 
\begin{align}
&\mathcal{E}_{\text{photon}}(\hat{\vec{n}}) = \int \widetilde{\mathrm{d}^{3} p} \, \omega_{p} \delta^2(z_{\hat{\vec{n}}} - z_{\hat{\vec{p}}}) \sum_{\sigma = \pm 1}: \Big[a^{\dagger}_{\sigma}(\omega_{p} \hat{\vec{n}}) a_{\sigma}(\omega_{p} \hat{\vec{n}}) \Big] :, \nonumber \\
&\mathcal{K}_{\text{photon}}(\hat{\vec{n}}) = \frac{ (-i)}{4} \int_{-\infty}^{+\infty}  \frac{\mathrm{d} \omega_{-}}{(2 \pi)^3} \delta^{(1)}(\omega_{-}) \int_{\text{max}\{-\omega_-, \omega_-\}}^{+\infty} \mathrm{d} \omega_{+} \, ((\omega_+)^2 - (\omega_-)^2)  \nonumber \\
& \hspace{100pt} \times : \sum_{\sigma = \pm 1} \Big[a^{\dagger}_{\sigma}\left((\omega_+ + \omega_-)\hat{\vec{n}}\right) a_{\sigma}\left((\omega_+ - \omega_-)\hat{\vec{n}}\right) \Big] : ,
\label{eqn:light_ray_gluon_EK}
\end{align}
because it is proportional to the number operator which counts the difference between the number of helicity plus and helicity minus photons
\begin{align}
\mathcal{N}^{\text{spin}}_{z,\text{photon}}(\hat{\vec{n}}) \ket{\{\vec{p},\sigma\}} = i \partial_{z_{\hat{\vec{n}}}}\delta^2(z_{\hat{\vec{n}}} - z_{\hat{\vec{p}}}) (\delta_{+,\sigma} - \delta_{-,\sigma}) \ket{\{\vec{p},\sigma\}}.
\end{align}
The commutation relation of $\mathcal{N}_z^\text{orb}$ with $\mathcal{N}_{\bar{z}}^\text{orb}$ is exactly the same as the scalar case
\begin{align}\label{eq:orb-orb}
&[\mathcal{N}^\text{orb}_{z,\text{photon}}(\hat{\vec{n}}_1), \mathcal{N}^\text{orb}_{\bar{z},\text{photon}}(\hat{\vec{n}}_2)] 
\nonumber \\ 
&\quad = 
	+ 2 i \partial_{z}  
		\delta^2(z_{\hat{\vec{n}}_1} - z_{\hat{\vec{n}}_2}) 
			\mathcal{N}_{\bar{z},\text{photon}}^\text{orb} (\hat{\vec{n}}_2)
	- 2 i \partial_{\bar{z}}  
		\delta^2( z_{\hat{\vec{n}}_1} - z_{\hat{\vec{n}}_2}) 	
			\mathcal{N}_{z,\text{photon}}^\text{orb} (\hat{\vec{n}}_2 )
	\nonumber \\
	&\qquad 
	-2 i \delta^2(z_{\hat{\vec{n}}_1} - z_{\hat{\vec{n}}_2}) 
		\partial_{z} 
			\mathcal{N}_{\bar{z},\text{photon}}^\text{orb} (\hat{\vec{n}}_2) .
\end{align}
This is obvious from the fact that the $\mathcal{N}_{z,\text{photon}}^\text{orb}$ is simply the sum of two copies of $\mathcal{N}_{z,\text{scalar}}^\text{orb}$
\begin{align} 
\mathcal{N}^{\text{orb}}_{z,\text{photon}}(\hat{\vec{n}}) = i\int \widetilde{\mathrm{d}^{3} p} \, \delta^2(z_{\hat{\vec{n}}} - z_{\hat{\vec{p}}})  \, \sum_{\sigma = \pm 1}:  \Big[a^{\dagger}_{\sigma}(\omega_{p} \hat{\vec{n}}) \overset\leftrightarrow{\partial}_{z_{\hat{\vec{n}}}} a_{\sigma}(\omega_{p} \hat{\vec{n}}) \Big] : .
\end{align}
Moreover it is clear that $\mathcal{N}_{z,\text{photon}}^\text{spin}(\hat{\vec{n}}_1)$ commutes with $\mathcal{N}_{\bar{z},\text{photon}}^\text{spin}(\hat{\vec{n}}_2)$
\begin{align} 
[\mathcal{N}_{z,\text{photon}}^\text{spin}(\hat{\vec{n}}_1), \mathcal{N}_{\bar{z},\text{photon}}^\text{spin}(\hat{\vec{n}}_2)\Big] = 0 .
\end{align}
The remaining orbital-spin commutators take the form
\begin{align} 
[\mathcal{N}_{z,\text{photon}}^\text{orb}(\hat{\vec{n}}_1), \mathcal{N}_{\bar{z},\text{photon}}^\text{spin}(\hat{\vec{n}}_2)\Big] &= 
	+ 2 i \partial_{z}  
		\delta^2( z_{\hat{\vec{n}}_1} - z_{\hat{\vec{n}}_2}) 	
			\mathcal{N}_{\bar{z},\text{photon}}^\text{spin} (\hat{\vec{n}}_2), \nonumber \\
[\mathcal{N}_{z,\text{photon}}^\text{spin}(\hat{\vec{n}}_1), \mathcal{N}_{\bar{z},\text{photon}}^\text{orb}(\hat{\vec{n}}_2)\Big] &=- 2 i \partial_{\bar{z}}  
		\delta^2( z_{\hat{\vec{n}}_1} - z_{\hat{\vec{n}}_2}) 	
			\mathcal{N}_{z,\text{photon}}^\text{spin} (\hat{\vec{n}}_2)
			\nonumber \\
			& \hspace{50pt}+ 2 i \delta^2(z_{\hat{\vec{n}}_1} - z_{\hat{\vec{n}}_2}) 
		\partial_{\bar{z}} \mathcal{N}_{z,\text{photon}}^\text{spin} (\hat{\vec{n}}_2) .
\end{align}
The above commutators are quite a bit more subtle than the others encountered so far and we do find an ambiguity in their derivation.\footnote{
Since the $\mathcal{N}_{z/\bar{z}}$ do not contain integrals over the transverse directions, one must smear them with a test function. A more rigorous analysis would involve studying Wightman functions in position space with a careful prescription for sending the operator positions to the same light-sheet (see \cite{Besken:2020snx}, for example) in addition to smearing with test functions.
} 
However, any problems with the orbital-spin commutator disappear in the integrated charges since $\mathcal{N}^\text{spin}_{z\slash\bar{z}}$ are total derivatives of $z$ or $\bar{z}$. 

If we turn on matter contributions, we find an additional (hard) interaction term which is mixing Coulombic and radiative data
\begin{align}
\int \mathrm{d} u \, A_{z}^{(0),a} j_u^{a,(2)} ,
\end{align}
which clearly breaks the algebra because it does not conserve the particle number. 

Regarding the soft non-abelian contribution which comes from the boundary term, we stress that as in the spin 0 case we are only interested in the hard contributions and therefore ignore the soft terms in our analysis, which requires a more delicate study of the soft non-abelian sector and its quantization as done by He and Mitra in  \cite{He:2020ifr}. We expect such terms to be related with the gluon soft theorem  \cite{Strominger:2013lka}, and therefore to be relevant for the light-ray algebra of pure Yang-Mills in the soft sector: the non-commutativity of soft gluon limits for different helicities might give a non-trivial extension of Cordova-Shao algebra. We leave this very interesting problem for a future study.

\subsection{Spin 2}

Here similar comments to the ones made in the spin 1 abelian case apply, except for the fact that we have some freedom in the choice of the Bondi angular momentum aspect and therefore of the density of the angular momentum flux as discussed before. We find that with our definitions the algebra of light-ray operators for the graviton case is consistent with the complexified Cordova-Shao algebra. In particular, there are no contributions which are mixing coulombic and radiative data as expected  \cite{Bonga:2018gzr}. Moreover, the convention adopted by Pasterski-Strominger-Zhiboedev  \cite{Pasterski:2015tva} seems to be the most natural for the system of light-ray operators since it provides the standard helicity term which we would expect from a spin 2 point particle stress tensor.

\subsection{Comparison with complexified Cordova-Shao algebra}
\label{sec:Cordova-Shao}

It is straightforward to check that the commutators 
\begin{align}
&[\mathcal{E}(\hat{\vec{n}}_1), \mathcal{E}(\hat{\vec{n}}_2)] = 0 ,\nonumber \\
&[\mathcal{K}(\hat{\vec{n}}_1), \mathcal{K}(\hat{\vec{n}}_2)] = 0 ,\nonumber \\
&[\mathcal{K}(\hat{\vec{n}}_1), \mathcal{E}(\hat{\vec{n}}_2)] = -2 i  \delta^2(z_{\hat{\vec{n}}_1} - z_{\hat{\vec{n}}_2}) \mathcal{E}(\hat{\vec{n}}_2) 
\end{align}
agree exactly with \cite{Cordova:2018ygx} since $T_{u u} = T_{- -}$\footnote{It is worth remembering here that $\delta^2(\Omega_{\hat{\vec{n}}_1} - \Omega_{\hat{\vec{n}}_2}) = 2 \delta^2(z_{\hat{\vec{n}}_1} - z_{\hat{\vec{n}}_2})$.}. However, in order to compare the other commutators, we need to compare how the components of the stress tensor at null infinity in our cooordinates differ from those in the standard light-cone components
\begin{align}
T_{u z} &= \frac{x^+}{2} \left(T_{- 1} - i T_{- 2} \right) + (x^1 - i x^2) \, T_{- -} ,  
\\
T_{u \bar{z}} &= \frac{x^+}{2} \left(T_{- 1} + i T_{- 2} \right) + (x^1 + i x^2) \, T_{- -} .
\label{eqn:Tcomponents}
\end{align}
Therefore, we see that a full detailed comparison with the (complexified) Cordova Shao algebra for the angular momentum flux requires understanding the following operator
\begin{align}
\lim_{x^+ \to +\infty} (x^+)^3 \, \left[\frac{1}{2} \left(T_{- 1} \pm i T_{- 2}\right)  + \frac{x^1 \pm i x^2}{x^+} T_{- -} \right] ,
\label{eqn:subleadingANEC}
\end{align}
where there is an explicit mixing with the original contribution of the complexified version of $\mathcal{N}_A$ at infinity. In general we have ($r = x^+$)
\begin{align}
 \lim_{r \to +\infty} r^2 T_{u u} &= \lim_{x^+ \to +\infty} (x^+)^2 T_{- -} ,\nonumber \\
\lim_{r \to +\infty} r^2 T_{u z} &=  \lim_{x^+ \to +\infty} (x^+)^3 \left[\frac{1}{2} \left(T_{- 1} - i T_{- 2} \right)  +  \frac{x^1 - i x^2}{x^+} \, T_{- -} \right] ,\nonumber \\
\lim_{r \to +\infty} r^2 T_{u \bar{z}} &= \lim_{x^+ \to +\infty} (x^+)^3 \left[ \frac{1}{2} \left(T_{- 1} + i T_{- 2} \right) + \frac{x^1 + i x^2}{x^+} \, T_{- -} \right] ,
\label{eqn:mixing}
\end{align}
which makes it clear how the operators are mixing with each other. Nevertheless, the commutators 
\begin{align}
&[\mathcal{N}_{z}(\hat{\vec{n}}_1), \mathcal{E}(\hat{\vec{n}}_2)] = -2 i \delta^2(z_{\hat{\vec{n}}_1} - z_{\hat{\vec{n}}_2}) \partial_{z}\mathcal{E}(\hat{\vec{n}}_2) + 2 i \partial_{z} \delta^2(z_{\hat{\vec{n}}_1} - z_{\hat{\vec{n}}_2}) \mathcal{E}(\hat{\vec{n}}_2) , \nonumber \\
& [\mathcal{N}_{z}(\hat{\vec{n}}_1), \mathcal{K}(\hat{\vec{n}}_2)] = -2 i \delta^2(z_{\hat{\vec{n}}_1} - z_{\hat{\vec{n}}_2}) \partial_{z}\mathcal{K}(\hat{\vec{n}}_2) + 2 i \partial_{z} \delta^2(z_{\hat{\vec{n}}_1} - z_{\hat{\vec{n}}_2}) \mathcal{K}(\hat{\vec{n}}_2)
\end{align}
match the complexified Cordova-Shao algebra. The only commutation relation which is new and differ from their result is $[\mathcal{N}_{z}(\hat{\vec{n}}_1), \mathcal{N}_{\bar{z}}(\hat{\vec{n}}_2)]$ \footnote{Their prediction is
\begin{align}
[\mathcal{N}_{z}(\hat{\vec{n}}_1), \mathcal{N}_{\bar{z}}(\hat{\vec{n}}_2)] &= 
	+ 2 i \partial_{z}  
		\delta^2(z_{\hat{\vec{n}}_1} - z_{\hat{\vec{n}}_2}) 
			\mathcal{N}_{\bar{z}} (\hat{\vec{n}}_2)
	+ 2 i \partial_{\bar{z}}  
		\delta^2( z_{\hat{\vec{n}}_1} - z_{\hat{\vec{n}}_2}) 	
			\mathcal{N}_{z} (\hat{\vec{n}}_2 ) -2 i \delta^2(z_{\hat{\vec{n}}_1} - z_{\hat{\vec{n}}_2}) \partial_{z} \mathcal{N}_{\bar{z}} (\hat{\vec{n}}_2) .
\nonumber
\end{align}
}, due to mixing in \eqref{eqn:mixing}.

\section{Gravitational energy event shapes and S-matrix}

In analogy to the standard event shapes considered in QCD jet physics, we explore gravitational event shapes, which are of current interest to the gravitational waves community. Our light-ray operators probe the underlying structure of gravitational radiation as registered by a detector placed at null infinity in the direction $\hat{\vec{n}}$. Of particular interest is the gravitational radiation emitted by the scattering of compact objects. 

We consider gravity as an effective field theory valid below the Planck scale  \cite{Donoghue:1994dn,Donoghue:2017pgk} and work in the weak field regime where we can rely on perturbation theory. We assume, for simplicity, that such compact objects are spinless and point-like; that is, they can be described massive scalar point particles of mass $m_1$ and $m_2$ minimally coupled to gravity. The relevant amplitude for such processes is of the form
\begin{align}
\ket{\text{in}} = \ket{\vec{p}_1 \vec{p}_2} \to \ket{\text{out}} = \ket{\vec{p}_3 \vec{p}_4  \{\vec{k}_j,\sigma_j\}_{j=1,..,M}} ,
\end{align}
where $\ket{\{\vec{k}_j,\sigma_j\}_{j=1,..,M}}$ are the graviton momentum eigenstates and $\ket{\vec{p}_i}$ are the momentum eigenstates of the two massive point particles.

In general, every radiation state in QFT can be expressed in terms of coherent states $\ket{\alpha}$ which provide an overcomplete basis  \cite{Glauber:1963tx,Glauber:1963fi}. Moreover in the case where the classical limit is well-defined the so-called Glauber-Sudarshan P-representation  \cite{Glauber:1963tx,Sudarshan:1963ts} of the radiation density matrix
\begin{align}
\hat{\rho}_{\text{radiation}} = \int P(\alpha) \ket{\alpha} \bra{\alpha} \mathrm{d}^2 \alpha 
\end{align}
admits only positive weights $P(\alpha) \geq 0$. In particular in suitable circumstances for soft radiation  \cite{Addazi:2019mjh,Cristofoli:2021vyo} or in the eikonal limit  \cite{Kabat:1992tb,Ciafaloni:2018uwe,Monteiro:2020plf} a single coherent state can represent the classical radiation at the quantum level.

Therefore, as a first approximation, we focus on the soft kinematic region of the emitted gravitons in order to take advantage of the simple form of graviton soft factors and of the fact the quantum state representing the radiation is known. To be more precise, the detectors are assumed to have a lower energy resolution $\lambda$ and the emitted gravitons to have total energy $E_{\text{wave}}$ with
\begin{align}
\lambda \ll E_{\text{wave}} \ll \Lambda ,
\end{align}
where $\Lambda$ is a suitable UV cutoff which can be thought as the total energy of the process \cite{Weinberg:1965nx,Addazi:2019mjh}. The classical limit requires the number of soft gravitons emitted to be large $M \to +\infty$, as dictated by the coherent state structure \cite{Laddha:2018rle}: this implies that the limit is quite delicate in general, and it implies an infinite resummation over all graviton contributions. 

Please note that we use Bondi retarded coordinates instead of flat-null coordinates in this section, in order to express everything in terms of the standard energy weight factor.

\subsection{The 1-point gravitational energy event shape}

The 1-point gravitational energy event shape is simply the on-shell expectation value of the gravitational ANEC at infinity
\begin{align}
\bra{\text{out}} \mathcal{\tilde{E}}(\hat{\vec{n}}) \ket{\text{out}} = \bra{\text{in}} S^{\dagger} \mathcal{\tilde{E}}(\hat{\vec{n}}) S \ket{\text{in}} .
\end{align}
Inserting the completeness relation of on-shell states $\sum_{X} \ket{\vec{X}} \bra{\vec{X}}$ yields
\begin{align}
\bra{\text{out}} \mathcal{\tilde{E}}(\hat{\vec{n}}) \ket{\text{out}} &= \sum_X \bra{\text{in}} S^{\dagger}  \mathcal{\tilde{E}}(\hat{\vec{n}}) \ket{\vec{X}} \bra{\vec{X}} S \ket{\text{in}} = \sum_X w_{ \mathcal{\tilde{E}}}(\hat{\vec{n}}) |\bra{\vec{X}} S \ket{\text{in}}|^2 ,
\end{align}
which is always positive definite.

The graviton emission amplitude for $M$ soft gravitons of momenta $k_j$ and helicity $\sigma_j$ is then\footnote{At leading order there is no distinction between consecutive double soft limits and simultaneous double soft limits, so we don't need to assume any hierarchy in the energy of the soft gravitons.}
\begin{align}
\hspace{-5pt}\mathcal{A}^{\text{GR}}(p_1,p_2;p_3,p_4,\{k_j,\sigma_j\}_{j=1,..,M}) &= \mathcal{A}^{\text{GR}}(p_1,p_2;p_3,p_4) \prod_{l = 1}^M \left[ \frac{\kappa}{2} \sum_{i = 1}^4 \eta_i   \frac{\epsilon^{\sigma_l}_{\mu \nu}(\vec{k}_l) p_i^{\mu} p_i^{\nu}}{p_i \cdot k_l} \right] ,
\label{eqn:soft_emission_amplitude}
\end{align}
where we have used the leading soft graviton factor  \cite{Weinberg:1964ew,Weinberg:1965nx} and $\eta_i = +1$ (resp. $\eta_i = -1$) if the particle is outgoing (resp. ingoing). One can go further in the soft expansion by taking into account the sub- (and subsub-) leading contributions, see   \cite{Sahoo:2018lxl,Laddha:2017ygw,Laddha:2018myi,Sen:2017nim,Cachazo:2014fwa} for further details.

We will be interested in the classical limit of this quantity, and the proper way to do so would be to use the in-in KMOC formalism  \cite{Kosower:2018adc}. In such approach, the limit is taken with a careful analysis of the wavepackets of the external massive fields which localize the particles on their classical trajectory as $\hbar \to 0$. Here for simplicity we will work in the approximation $\omega \ll b^{-1}$, where $\omega$ is the typical graviton frequency of the emitted wave and $b$ is the impact parameter of the two incoming particles: this will allow us to get a simple universal $b$-independent result which will be useful to analyze the properties of gravitational energy event shapes. In order to make transparent the power counting for the amplitudes involved, we will consider the tree level classical contribution for $\mathcal{A}^{\text{GR}}(p_1,p_2;p_3,p_4)$ so that $|\mathcal{A}^{\text{GR}}(p_1,p_2;p_3,p_4)|^2$ will always be of order $G^2$ in this section. 

The leading contribution will be given by the single graviton emission amplitude,
\begin{align}
& \bra{\text{out}} \mathcal{\tilde{E}}_{\text{GR}}(\hat{\vec{n}}) \ket{\text{out}} \Big|_{\mathcal{O}(G^{3})} =  \int \widetilde{\mathrm{d}^{3} p_3} \int \widetilde{\mathrm{d}^{3} p_4} \sum_{\sigma_1 = \pm 2} \int \widetilde{\mathrm{d}^{3} k_1} \nonumber \\
&\times \left[(E_{k_1}) \delta^2(\Omega_{\vec{\hat{k}}_1} - \Omega_{\hat{\vec{n}}}) \right] (2 \pi)^4 \delta^4\left(\sum_{i=1}^4 \eta_i p_i + k_1\right) |\mathcal{A}^{\text{GR}}(p_1,p_2;p_3,p_4,\{k_1,\sigma_1\})|^2 .
\end{align}
In the soft regime for the graviton we can isolate the graviton phase space integration,
\begin{align}
& \sum_{\sigma_1 = \pm 2} \int \widetilde{\mathrm{d}^{3} k_1}  (2 \pi)^4 \delta^4\left(\sum_{i=1}^4 \eta_i  p_i + k_1\right) \left[ (E_{k_1}) \delta^2(\Omega_{\vec{\hat{k}}_1} - \Omega_{\hat{\vec{n}}}) \right] |\mathcal{A}^{\text{GR}}(p_1,p_2;p_3,p_4,\{k_1,\sigma_1\})|^2 \nonumber \\
&\simeq (2 \pi)^4 \delta^4\left(\sum_{i=1}^4 \eta_i  p_i\right) \sum_{\sigma_1 = \pm 2} \int \widetilde{\mathrm{d}^{3} k_1} \left[(E_{k_1}) \delta^2(\Omega_{\vec{\hat{k}}_1} - \Omega_{\hat{\vec{n}}}) \right]  |\mathcal{A}^{\text{GR}}(p_1,p_2;p_3,p_4,\{k_1,\sigma_1\})|^2 .
\end{align}
To make further progress we need to use the polarization sum identity for the graviton polarization vectors which reads
\begin{align}
\sum_{\sigma = \pm 2} \epsilon^{\sigma}_{\mu \nu}(k) \epsilon^{\sigma, *}_{\alpha \beta}(k) &= \frac{1}{2} \left[P_{\mu \alpha} P_{\nu \beta} + P_{\mu \beta} P_{\nu \alpha} - P_{\mu \nu} P_{\alpha \beta} \right] , \nonumber \\
P_{\mu \nu} &= \eta_{\mu \nu} - \frac{q_{\mu} p_{\nu} + q_{\nu} p_{\mu}
}{p \cdot q} ,
\end{align}
where $q^{\mu}$ is a suitable reference momentum. The gauge dependent terms will cancel from our calculation using momentum conservation   \cite{Weinberg:1965nx}, so we do not need to worry about spurious contributions. Since the amplitude factorizes, we can actually perform the graviton phase space integration in a universal form
\begin{align}
&\sum_{\sigma_1 = \pm 2} \frac{1}{2 (2 \pi)^3} \int_{\lambda}^{E_{\text{wave}}} \mathrm{d} E_{k_1} \int \mathrm{d} \Omega_{\vec{\hat{k}}_1} \,\delta^2(\Omega_{\vec{\hat{k}}_1} - \Omega_{\hat{\vec{n}}}) (E_{k_1})^2 \Bigg|\sum_{i = 1}^4 \eta_i \frac{\kappa}{2} \frac{\epsilon^{\sigma_1}_{\mu \nu}(\vec{k}_1) p_i^{\mu} p_i^{\nu}}{p_i \cdot k_1} \Bigg|^2  \nonumber \\
&= G \frac{E_{\text{wave}}}{ 4 \pi^2} \left(\sum_{i,j = 1}^4  \eta_i \eta_j \frac{2 (p_i \cdot p_j)^2 - m_i^2 m_j^2}{(E_{p_i} - \vec{p}_i \cdot \hat{\vec{n}})(E_{p_j} - \vec{p}_j \cdot \hat{\vec{n}})}\right) .
\end{align}
Working in the center of mass frame\footnote{Here $s=(p_1 + p_2)^2$ is the center of mass energy.} for the two massive particles we have
\begin{align}
& \bra{\text{out}} \mathcal{\tilde{E}}_{\text{GR}}(\hat{\vec{n}}) \ket{\text{out}} \Big|_{\mathcal{O}(G^{3})}= \int \mathrm{d} \Omega_{\vec{p}_{\text{CM}}} \, \frac{|\vec{p}_{\text{CM}}|}{16 \pi^2 \sqrt{s}} |\mathcal{A}^{\text{GR}}(p_1,p_2;p_3,p_4)|^2 \nonumber \\
&\hspace{130pt}\times G \frac{E_{\text{wave}} }{4 \pi^2} \left(\sum_{i,j = 1}^4  \eta_i \eta_j \frac{2 (p_i \cdot p_j)^2 - m_i^2 m_j^2}{(E_{p_i} - \vec{p}_i \cdot \hat{\vec{n}})(E_{p_j} - \vec{p}_j \cdot \hat{\vec{n}})}\right) .
\label{eqn:1-pt_energyk4}
\end{align}
which is infrared finite at the leading order $G^{3}$. It is worth noticing that \eqref{eqn:1-pt_energyk4} is only the leading order contribution, and we still need to sum over all possible graviton emissions.

\begin{figure}[h!]
  \centering
  \includegraphics[width=1\textwidth]{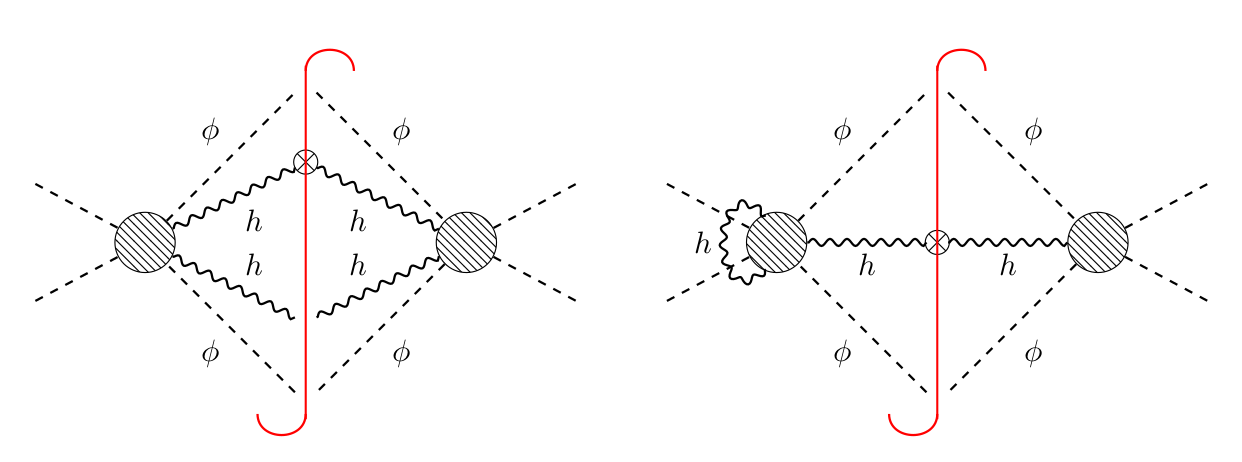}
\caption{The picture shows how the cancellation of infrared divergences between real and virtual graviton emissions works for the gravitational energy event shape} 
\label{fig:infrared_cancellation}
\end{figure}

Before looking at the classical limit, it is worth looking at how infrared divergences will cancel for the gravitational energy event shape in general. At order $G^{4}$ we have,
\begin{align}
& \bra{\text{out}} \mathcal{\tilde{E}}_{\text{GR}}(\hat{\vec{n}}) \ket{\text{out}} \Big|_{\mathcal{O}(G^{4})} \nonumber \\
&=  \int \widetilde{\mathrm{d}^{3} p_3} \int \widetilde{\mathrm{d}^{3} p_4} \sum_{\sigma_1 = \pm 2} \int \widetilde{\mathrm{d}^{3} k_1}  \, (2 \pi)^4 \delta^4\left(\sum_{i=1}^4 \eta_i p_i + k_1\right) \nonumber \\
&\hspace{50pt} \times \left[(E_{k_1}) \delta^2(\Omega_{\vec{\hat{k}}_1} - \Omega_{\hat{\vec{n}}}) \right]  |\mathcal{A}^{\text{GR}}(p_1,p_2;p_3,p_4,\{k_1,\sigma_1\})|^2 \nonumber \\
&+  \int \widetilde{\mathrm{d}^{3} p_3} \int \widetilde{\mathrm{d}^{3} p_4} \, \frac{1}{2}  \sum_{\sigma_1,\sigma_2 = \pm 2} \int \widetilde{\mathrm{d}^{3} k_1} \int \widetilde{\mathrm{d}^{3} k_2} \, (2 \pi)^4 \delta^4\left(\sum_{i=1}^4 \eta_i p_i + \sum_{l=1}^2 k_l\right)\nonumber \\
&\hspace{50pt} \times \left[\sum_{l = 1}^2 (E_{k_l}) \delta^2(\Omega_{\vec{\hat{k}}_l} - \Omega_{\hat{\vec{n}}}) \right]  |\mathcal{A}^{\text{GR}}(p_1,p_2;p_3,p_4,\{k_1,\sigma_1\},\{k_2,\sigma_2\})|^2 .
\label{eqn:1-pt_energyk6}
\end{align}
The first contribution includes loop contributions related to virtual gravitons 
\begin{align}
&|\mathcal{A}^{\text{GR}}(p_1,p_2;p_3,p_4,\{k_1,\sigma_1\})|^2\Big|_{\mathcal{O}(G^{4})} = |\mathcal{A}^{\text{GR}}(p_1,p_2;p_3,p_4,\{k_1,\sigma_1\})|^2\Big|_{\mathcal{O}(G^{3})}  \nonumber \\
& \times \left[- \frac{G}{4 \pi^2} \log\left(\frac{\Lambda}{\lambda}\right) \int \mathrm{d} \Omega_{\vec{\hat{k}}}  \left(\sum_{i,j = 1}^4  \eta_i \eta_j  \frac{2 (p_i \cdot p_j)^2 -  m_i^2 m_j^2}{ (E_{p_i} - \vec{p}_i \cdot \hat{\vec{k}})(E_{p_j} - \vec{p}_j \cdot \hat{\vec{k}})}\right)  \right] , 
\end{align}
where we have regularized the integral with the ultraviolet cutoff $\Lambda$ \cite{Weinberg:1965nx}. The second term instead has a divergent contribution which comes from the emission of one real graviton, since integrating over the 2-gravitons phase space gives
\begin{align}
& \frac{1}{2} \sum_{\sigma_1,\sigma_2 = \pm 2} \int_{\lambda}^{E_{\text{wave}}} \frac{\mathrm{d} E_{k_1} (E_{k_1})}{2 (2 \pi)^3} \int_{\lambda}^{E_{\text{wave}}} \frac{\mathrm{d} E_{k_2} (E_{k_2})}{2 (2 \pi)^3} \int \mathrm{d} \Omega_{\vec{\hat{k}}_1} \, \int \mathrm{d} \Omega_{\vec{\hat{k}}_2} \, \Theta(E_{\text{wave}} - E_{k_1} - E_{k_2}) \nonumber \\
& \hspace{70pt} \times \left[\sum_{l=1}^2(E_{k_l}) \delta^2(\Omega_{\vec{\hat{k}}_l} - \Omega_{\hat{\vec{n}}}) \right] \prod_{l=1}^2\Bigg|\sum_{i = 1}^4 \eta_i \frac{\kappa}{2} \frac{\epsilon^{\sigma_l}_{\mu \nu}(\vec{k}_l) p_i^{\mu} p_i^{\nu}}{p_i \cdot k_l} \Bigg|^2  \nonumber \\
&= G \frac{E_{\text{wave}}}{4 \pi^2} \left(\sum_{i,j = 1}^4  \eta_i \eta_j \frac{2 (p_i \cdot p_j)^2 - m_i^2 m_j^2}{(E_{p_i} - \vec{p}_i \cdot \hat{\vec{n}})(E_{p_j} - \vec{p}_j \cdot \hat{\vec{n}})}\right) \nonumber \\
& \hspace{70pt} \times \log\left(\frac{E_{\text{wave}}}{\lambda}\right) \frac{G}{4 \pi^2}  \left[\int \mathrm{d} \Omega_{\vec{\hat{k}}} \sum_{i,j = 1}^4  \eta_i \eta_j \frac{2 (p_i \cdot p_j)^2 - m_i^2 m_j^2}{(E_{p_i} - \vec{p}_i \cdot \hat{\vec{k}})(E_{p_j} - \vec{p}_j \cdot \hat{\vec{k}})}\right] ,
\end{align}
where we have used the fact that
\begin{align}
\int_{\lambda}^{E_{\text{wave}}} \mathrm{d} E_{k_1} \, \int_{\lambda}^{E_{\text{wave}}} \mathrm{d} E_{k_2} \, \frac{1}{E_{k_2}}\Theta(E_{\text{wave}} - E_{k_1} - E_{k_2}) \stackrel{E_{k_1},E_{k_2} \ll E_{\text{wave}}}{\simeq} E_{\text{wave}}  \log\left(\frac{E_{\text{wave}}}{\lambda}\right) ,
\end{align}
which is valid as long as we can neglect the single graviton energies (suppressed by $\hbar$) compared to the energy of the wave $E_{k_1},E_{k_2} \ll E_{\text{wave}}$, as noticed in  \cite{Addazi:2019mjh}.

One can then perform the regularized sum in \eqref{eqn:1-pt_energyk6}
\begin{align}
& \hspace{-15pt}\bra{\text{out}} \mathcal{\tilde{E}}_{\text{GR}}(\hat{\vec{n}}) \ket{\text{out}} \Big|_{\mathcal{O}(G^{4})} =   \int \mathrm{d} \Omega_{\vec{p}_{\text{CM}}} \frac{|\vec{p}_{\text{CM}}|}{16 \pi^2 \sqrt{s}} |\mathcal{A}^{\text{GR}}(p_1,p_2;p_3,p_4)|^2 \nonumber \\
& \times G \frac{E_{\text{wave}}}{4 \pi^2}  \left(\sum_{i,j = 1}^4  \eta_i \eta_j \frac{2 (p_i \cdot p_j)^2 - m_i^2 m_j^2}{(E_{p_i} - \vec{p}_i \cdot \hat{\vec{n}})(E_{p_j} - \vec{p}_j \cdot \hat{\vec{n}})}\right)  \nonumber \\
& \times \log\left(\frac{E_{\text{wave}}}{\Lambda}\right) \frac{G}{4 \pi^2} \left[\int \mathrm{d} \Omega_{\vec{\hat{k}}}  \sum_{i,j = 1}^4  \eta_i \eta_j \frac{2 (p_i \cdot p_j)^2 - m_i^2 m_j^2}{(E_{p_i} - \vec{p}_i \cdot \hat{\vec{k}})(E_{p_j} - \vec{p}_j \cdot \hat{\vec{k}})}\right] ,
\end{align}
which is infrared finite as expected. One can then repeat the same argument for the $G^{3 + n}$-th order contribution
\begin{align}
& \hspace{-15pt}\bra{\text{out}} \mathcal{E}_{\text{GR}}(\hat{\vec{n}}) \ket{\text{out}} \Big|_{\mathcal{O}(G^{3 + n})} =  \int \mathrm{d} \Omega_{\vec{p}_{\text{CM}}} \frac{|\vec{p}_{\text{CM}}|}{16 \pi^2 \sqrt{s}} |\mathcal{A}^{\text{GR}}(p_1,p_2;p_3,p_4)|^2 \nonumber \\
& \times G \frac{E_{\text{wave}} }{4 \pi^2} \left(\sum_{i,j = 1}^4  \eta_i \eta_j \frac{2 (p_i \cdot p_j)^2 - m_i^2 m_j^2}{(E_{p_i} - \vec{p}_i \cdot \hat{\vec{n}})(E_{p_j} - \vec{p}_j \cdot \hat{\vec{n}})}\right) \nonumber \\
& \times \sum_{M^{\prime} =0}^n \frac{1}{M^{\prime}!} \log\left(\frac{E_{\text{wave}}}{\Lambda}\right)^{M^{\prime}} \left[\frac{G}{4 \pi^2}\int \mathrm{d} \Omega_{\vec{\hat{k}}}  \sum_{i,j = 1}^4  \eta_i \eta_j \frac{2 (p_i \cdot p_j)^2 - m_i^2 m_j^2}{(E_{p_i} - \vec{p}_i \cdot \hat{\vec{k}})(E_{p_j} - \vec{p}_j \cdot \hat{\vec{k}})}\right]^{M^{\prime}} ,
\label{eqn:1-pt_energykn}
\end{align}
where $M^{\prime} = M - 1$.

It is worth noticing that this situation is in contrast with QCD case where the gluon energy event shape is infrared divergent and one has to sum also over quarks energy event shapes to get a well-defined (i.e. an infrared finite) answer  \cite{Basham:1977iq}. The cancellation of infrared divergences in this setting is similar to what was found for the graviton emission cross section by Donoghue  \cite{Donoghue:1999qh}. 

\subsection{The 2-points gravitational energy event shape}

The main object of interest here will be 
\begin{align}
\bra{\text{out}} \mathcal{\tilde{E}}(\hat{\vec{n}}_1) \mathcal{\tilde{E}}^{\prime}(\hat{\vec{n}}_2) \ket{\text{out}} = \bra{\text{in}} S^{\dagger} \mathcal{\tilde{E}}(\hat{\vec{n}}_1) \mathcal{\tilde{E}}^{\prime}(\hat{\vec{n}}_2) S \ket{\text{in}} .
\end{align}
By inserting a completeness relation of on-shell final states $\sum_{X} \ket{\vec{X}} \bra{\vec{X}}$ between the two gravitational ANEC operators at infinity we get
\begin{align}
\bra{\text{out}} \mathcal{\tilde{E}}(\hat{\vec{n}}_1) \mathcal{\tilde{E}}^{\prime}(\hat{\vec{n}}_2) \ket{\text{out}} &= \sum_X \bra{\text{in}} S^{\dagger} \mathcal{\tilde{E}}(\hat{\vec{n}}_1) \ket{\vec{X}} \bra{\vec{X}} \mathcal{\tilde{E}}^{\prime}(\hat{\vec{n}}_2) S \ket{\text{in}} \nonumber \\
&= \sum_X w_{\mathcal{\tilde{E}}}(\hat{\vec{n}}_1) w^{\prime}_{\mathcal{\tilde{E}}}(\hat{\vec{n}}_2) |\bra{\vec{X}} S \ket{\text{in}}|^2 .
\end{align}
In details
\begin{align}
&\hspace{-5pt}\bra{\text{out}} \mathcal{\tilde{E}}_{\text{GR}}(\hat{\vec{n}}_1) \mathcal{\tilde{E}}^{\prime}_{\text{GR}}(\hat{\vec{n}}_2) \ket{\text{out}} \Big|_{\mathcal{O}(G^{4})} \nonumber \\
&\hspace{-5pt} = \int \widetilde{\mathrm{d}^{3} p_3} \int \widetilde{\mathrm{d}^{3} p_4} \frac{1}{2} \sum_{\sigma_1,\sigma_2 = \pm 2} \int \widetilde{\mathrm{d}^{3} k_1} \int \widetilde{\mathrm{d}^{3} k_2}   \nonumber \\
&\hspace{-5pt}\times \Big[(E_{k_1})(E_{k_1}) \delta^2(\Omega_{\vec{\hat{k}}_1} - \Omega_{\hat{\vec{n}}_1}) \delta^2(\Omega_{\vec{\hat{k}}_1} - \Omega_{\hat{\vec{n}}_2}) + (E_{k_2})(E_{k_2}) \delta^2(\Omega_{\vec{\hat{k}}_2} - \Omega_{\hat{\vec{n}}_1}) \delta^2(\Omega_{\vec{\hat{k}}_2} - \Omega_{\hat{\vec{n}}_2}) \nonumber \\
&\hspace{7pt} +2 (E_{k_1})(E_{k_2}) \delta^2(\Omega_{\vec{\hat{k}}_1} - \Omega_{\hat{\vec{n}}_1}) \delta^2(\Omega_{\vec{\hat{k}}_2} - \Omega_{\hat{\vec{n}}_2})  \Big]|\bra{\{\vec{k}_1,\sigma_1\} \{\vec{k}_2,\sigma_2\} \vec{p}_3 \vec{p}_4} S \ket{\vec{p}_1 \vec{p}_2}|^2 .
\end{align}
Contrary to the 1-pt energy event shape, here we get also contact terms which correspond to the case where the two detectors are aligned along the same direction. These terms are usually included in QCD energy event shapes and they are required in order to remove collinear divergences \cite{Basham:1978bw}, but classical gravity does not have such type of collinear divergences  \cite{Akhoury:2011kq}. We will thus restrict our attention to the generic case $\hat{\vec{n}}_1 \neq \hat{\vec{n}}_2$:
\begin{align}
&\bra{\text{out}} \mathcal{\tilde{E}}_{\text{GR}}(\hat{\vec{n}}_1) \mathcal{\tilde{E}}^{\prime}_{\text{GR}}(\hat{\vec{n}}_2) \ket{\text{out}} \Big|_{\mathcal{O}(G^{4})} =  \int \widetilde{\mathrm{d}^{3} p_3} \int \widetilde{\mathrm{d}^{3} p_4} \,\frac{1}{2} \sum_{\sigma_1,\sigma_2 = \pm 2} \int \widetilde{\mathrm{d}^{3} k_1} \int \widetilde{\mathrm{d}^{3} k_2} \nonumber \\
& \hspace{50pt}\times ( 2 E_{k_1} E_{k_2} \delta^2(\Omega_{\vec{\hat{k}}_1} - \Omega_{\hat{\vec{n}}_1}) \delta^2(\Omega_{\vec{\hat{k}}_2} - \Omega_{\hat{\vec{n}}_2}) ) |\bra{\{\vec{k}_1,\sigma_1\} \{\vec{k}_2,\sigma_2\} \vec{p}_3 \vec{p}_4} S \ket{\vec{p}_1 \vec{p}_2}|^2 .
\end{align}
At this point we can repeat the integration as we did before to get
\begin{align}
&\bra{\text{out}} \mathcal{\tilde{E}}_{\text{GR}}(\hat{\vec{n}}_1) \mathcal{\tilde{E}}^{\prime}_{\text{GR}}(\hat{\vec{n}}_2) \ket{\text{out}} \Big|_{\mathcal{O}(G^{4})} = \int \mathrm{d} \Omega_{\vec{p}_{\text{CM}}} \frac{|\vec{p}_{\text{CM}}|}{16 \pi^2 \sqrt{s}} |\mathcal{A}^{\text{GR}}(p_1,p_2;p_3,p_4)|^2  \nonumber \\
& \hspace{60pt} \times \prod_{l=1}^2 \left[G \frac{E_{\text{wave}}}{4 \pi^2} \left(\sum_{i,j = 1}^4  \eta_i \eta_j \frac{2 (p_i \cdot p_j)^2 - m_i^2 m_j^2}{(E_{p_i} - \vec{p}_i \cdot \hat{\vec{n}}_l)(E_{p_j} - \vec{p}_j \cdot \hat{\vec{n}}_l)}\right) \right] .
\end{align}
As we can see, despite being a natural quantum observable the 2-point gravitational energy event shape factorizes, i.e. there is no degree of correlation between the two emissions. This is due to the soft  limit we are considering and ultimately related to the coherent state structure of the radiation  \cite{Ciafaloni:2018uwe}. It would be very interesting to compute the non-trivial (connected) 2-point energy event shape in a quantum theory of gravity from the six point amplitude with 4 massive scalars and 2 gravitons, which should provide the leading infrared finite contribution to this observable.

\subsection{Gravitational wave energy event shapes and classical factorization}

So far we have discussed the quantum picture, where energy event shape correlators are a non-trivial infrared-safe prediction. For our problem in the classical soft limit, as we will see, only the 1-point event shape will be non-trivial because of factorization properties. Let's first discuss how to achieve the classical limit. We are interested in the limit where those gravitons contribute coherently to a (suitably localized) wave travelling in a specific direction. To make this precise, one can study the interaction hamiltonian for our theory of scalars minimally coupled with gravity in the linearized gravity approximation and in asymptotic limit $|t| \to +\infty$  \cite{Ware:2013zja}
\begin{align}
H(t) &= H_0 + V^{\text{asy}} (t) = H_0 - \int \mathrm{d}^3 x \, h^{\mu \nu}(t,\vec{x}) T^{\text{asy}}_{\mu \nu}(t,\vec{x}) , \nonumber \\
 T^{\text{asy}}_{\mu \nu} &= \sum_{i=1,2} \frac{\kappa}{2} \int \widetilde{\mathrm{d}^{3} p} \, \frac{p^{\mu} p^{\nu}}{E_{p}} \rho_i(\vec{p}) \delta^3\left(\vec{x} - t \frac{\vec{p}}{E_{p}}\right) ,
\label{eqn:FK_Hamiltonian}
\end{align}
where $H_0$ is the free hamiltonian, $\rho_i(\vec{p}) = b^{\dagger}_i(\vec{p}) b_i(\vec{p})$ is the number operator for the $i$-th scalar field and $(t,\vec{x})$ are the coordinates of the asymptotic trajectory of the scalar particles. This result is is valid only at first order in the perturbation $h_{\mu \nu}$, but it tells something important about the asymptotic dynamics: solving the asymptotic Schr\"odinger equation for the potential $V^{\text{asy}}(t)$ gives an evolution operator which generates an infinite number of (soft) gravitons, or more precisely a coherent state of gravitons\footnote{There is also an additional term which is the analogue of the Coulomb phase contribution in QED, which plays an important role at higher orders in the soft expansion but not at leading order.}
\begin{align}
\ket{\{\alpha_p,\sigma\}} = N_{\alpha} \exp\left[ \int \widetilde{\mathrm{d}^{3} p} \left( \alpha_p(\vec{p}) a^{\dagger}_{\sigma}(\vec{p})-\mathrm{h.c.}\right)\right] \ket{0} 
\label{eqn:coherent_state}
\end{align}
for a suitable function $\alpha_p(\vec{p})$ related to the structure of the asymptotic stress momentum tensor of the scalar particle $T^{\text{asy}}_{\mu \nu}$. In this specific context the state \eqref{eqn:coherent_state} is referred as Faddeev-Kulish state  \cite{Kulish:1970ut}. This gives an insight about how the quantum scattering theory generates a classical gravitational wave, composed of infinitely many gravitons, at least in the soft kinematic regime. More details about the connection of coherent states with the classical limit will be provided in   \cite{Cristofoli:2021vyo}. 

In our case we can take the sum over all contributions $M = 1,..,+\infty$, which implies to sum over $M^{\prime} = 0,..,+\infty$ in our expression for the gravitational wave event shape in \eqref{eqn:1-pt_energykn}:
\begin{align}
&\hspace{-1pt} \bra{\text{out}} \mathcal{\tilde{E}}_{\text{GR}}(\hat{\vec{n}}) \ket{\text{out}} \Big|_{\text{class}} =  G \frac{E_{\text{wave}} }{4 \pi^2}  \int \mathrm{d} \Omega_{\vec{p}_{\text{CM}}} \frac{|\vec{p}_{\text{CM}}|}{16 \pi^2 \sqrt{s}} |\mathcal{A}^{\text{GR}}(p_1,p_2;p_3,p_4)|^2 \nonumber \\
& \times \left(\sum_{i,j = 1}^4  \eta_i \eta_j \frac{2 (p_i \cdot p_j)^2 - m_i^2 m_j^2}{(E_{p_i} - \vec{p}_i \cdot \hat{\vec{n}})(E_{p_j} - \vec{p}_j \cdot \hat{\vec{n}})}\right) \left(\frac{E_{\text{wave}}}{\Lambda}\right)^{\frac{G}{2 \pi} \sum_{i, j} \eta_i \eta_j \frac{m_i m_j (1 + \beta^2_{i j})}{\beta_{i j} (1 - \beta^2_{i j})^{\frac{1}{2}}} \log\left(\frac{1 + \beta_{i j}}{1 - \beta_{i j}}\right) } ,
\label{eqn:1-pt_energywave}
\end{align}
where $\beta_{i j} = \sqrt{1 - \frac{m_i^2 m_j^2}{(p_i \cdot p_j)^2}}$. This is consistent with the graviton production rate in Weinberg \cite{Weinberg:1965nx} under our hypotheses. Once normalized by the total cross section we have
\begin{align}
&\hspace{-1pt} \Big\langle \mathcal{\tilde{E}}_{\text{GR}}(\hat{\vec{n}}) \Big\rangle \Big|_{\text{class}} = \frac{E_{\text{wave}} G}{4 \pi^2} \left(\sum_{i,j = 1}^4  \eta_i \eta_j \frac{2 (p_i \cdot p_j)^2 - m_i^2 m_j^2}{(E_{p_i} - \vec{p}_i \cdot \hat{\vec{n}})(E_{p_j} - \vec{p}_j \cdot \hat{\vec{n}})}\right) \nonumber \\
&\hspace{160pt} \times \left(\frac{E_{\text{wave}}}{\Lambda}\right)^{\frac{G}{2 \pi} \sum_{i, j} \eta_i \eta_j \frac{m_i m_j (1 + \beta^2_{i j})}{\beta_{i j} (1 - \beta^2_{i j})^{\frac{1}{2}}} \log\left(\frac{1 + \beta_{i j}}{1 - \beta_{i j}}\right) }
\end{align}
and if we try to compute the gravitational energy-energy correlator at the classical level
\begin{align}
&\hspace{-1pt} \Big\langle  \mathcal{\tilde{E}}_{\text{GR}}(\hat{\vec{n}}_1) \mathcal{\tilde{E}}^{\prime}_{\text{GR}}(\hat{\vec{n}}_2)) \Big\rangle \Big|_{\text{class}} = \Big\langle \mathcal{\tilde{E}}_{\text{GR}}(\hat{\vec{n}}_1) \Big\rangle \Big|_{\text{class}} \Big\langle \mathcal{\tilde{E}}^{\prime}_{\text{GR}}(\hat{\vec{n}}_2) \Big\rangle \Big|_{\text{class}} ,
\end{align}
which is a consequence of the classical (soft) factorization. This can be shown explicitly using the coherent state in \eqref{eqn:coherent_state}. 

\subsection{Infrared safety of gravitational energy event shapes and soft BMS supertranslation}

It is possible to do an interesting consistency check about infrared safety of the gravitational energy event shapes by looking at how soft supertranslation symmetry affects such matrix elements.  Earlier we have used an infrared cutoff and we have observed the cancellation of infrared divergences as in the Bloch-Nordsieck mechanism, but it would be nice to understand if there is a theoretical reason for these event shapes to be infrared finite.

It is well known that the action of the soft supertranslation mode corresponds to an insertion of a soft graviton at the level of S-matrix elements  \cite{Strominger:2013jfa,He:2014laa}. In particular the generator of such symmetry reads 
\begin{align}
T_{\text{soft}}(f) &= \frac{1}{16 \pi G} \int \mathrm{d}^2 \zeta \left[\partial_{\zeta} D^{\bar{\zeta}} C_{\bar{\zeta} \bar{\zeta}} + \partial_{\bar{\zeta}} D^{\zeta} C_{\zeta \zeta} \right] f(\zeta,\bar{\zeta}) \nonumber \\
&= \lim_{E \to 0} \frac{E}{4 \pi \kappa} \int \mathrm{d}^2 \zeta \,\left[ (a_+(E \hat{\vec{n}}) + a^{\dagger}_-(E \hat{\vec{n}})) D_{\bar{\zeta}}^2 f + h.c. \right] ,
\end{align}
where $f(\zeta,\bar{\zeta})$ is an arbitrary function on the celestial sphere. Its action on graviton creation operators is given by
\begin{align}
[T_{\text{soft}}(f),a_+(\vec{k})] = \frac{8 \pi^2}{\kappa} \frac{\delta(E_k)}{\gamma_{\zeta_k \bar{\zeta}_k}} D^2_{\zeta} f \,, \qquad [T_{\text{soft}}(f),a_-(\vec{k})] = \frac{8 \pi^2}{\kappa} \frac{\delta(E_k)}{\gamma_{\zeta_k \bar{\zeta}_k}} D^2_{\bar{\zeta}} f ,
\label{eqn:commutation}
\end{align}
and the one with creation operators is fixed because $T_{\text{soft}}(f)$ is hermitian. The latter implies that 
\begin{align}
[T_{\text{soft}}(f),a^{\dagger}_{\sigma}(\vec{k})]^{\dagger} = - [T_{\text{soft}}(f),a_{\sigma}(\vec{k})] \qquad \forall \,\sigma=\pm 2 .
\end{align}
We now consider a gravitational event shape of the form
\begin{align}
\bra{\text{out}} \mathcal{\tilde{P}}_{\mu,\text{GR}}(\hat{\vec{n}}) \ket{\text{out}} = \sum_X \bra{\text{in}} S^{\dagger} \mathcal{\tilde{P}}_{\mu,\text{GR}}(\hat{\vec{n}}) \ket{\vec{X}} \bra{\vec{X}} S \ket{\text{in}}
\label{eqn:grav_event_shape}
\end{align}
and we want to check whether this definition is invariant or not under BMS soft supertranslations, i.e. under the addition of a soft graviton. If it is, this provides a strong evidence that \eqref{eqn:grav_event_shape} is actually infrared finite, i.e. insensitive to soft (graviton) physics. A short calculation shows that 
\begin{align}
[T_{\text{soft}}(f),\mathcal{\tilde{P}}_{\mu,\text{GR}}(\hat{\vec{n}})] = \frac{n_{\mu}}{2 \pi \kappa \gamma_{\zeta_{\hat{\vec{n}}} \bar{\zeta}_{\hat{\vec{n}}}}} \lim_{E_p \to 0} (E_p)^2 \left[D^2_{\zeta} f \, a^{\dagger}_{+}(E_p \hat{\vec{n}}) + D^2_{\bar{\zeta}} f \, a^{\dagger}_{-}(E_p \hat{\vec{n}}) - h.c.\right] 
\label{eqn:T_soft_action_P}
\end{align}
and since the addition of a soft graviton gives the usual Weinberg soft factor with a pole in $E_p$, the action \eqref{eqn:T_soft_action_P} annihilates every S-matrix element (once we insert the completeness relation)  \cite{He:2014laa}.

It would be interesting to understand in more details if the invariance under the soft modes of large gauge symmetries, BMS supertranslations and in general asymptotic symmetries of massless particles always guarantees the IR-finiteness (in the soft sense\footnote{Here we are not considering collinear divergences, which play a huge role in QCD.}) of a matrix element in a QFT. This could be helpful to prove in general whether for other on-shell observables, for example like the ones defined in   \cite{Kosower:2018adc}, infrared divergences are always going to cancel.

\section{Discussion and Conclusions} 

Einstein equations solved near null infinity and BMS balance laws are of particular relevance for studying the property of the radiation at infinity, like its energy and its angular momentum which are captured by detectors localized at spatial infinity in a particular direction $\vec{n}$. Once we integrate Einstein equations over the retarded time, say in Bondi gauge, new non-local operators appear which are commonly called light-ray operators. For our definition of light-ray operators, we integrate over the entire retarded time interval $u \in \mathbb{R}$, which allows us to extract a suitable notion of density of energy and angular momentum flux produced during a scattering event. This is the natural framework where contributions coming from different massless particles combine together in a single expression. 

Besides the direct physical applications, the algebra of such non-local light-ray operators at infinity is of intrinsic interest since it can be considered some sort of ``detector algebra''. In the conformal case the algebra was clarified in a beautiful work by Cordova and Shao  \cite{Cordova:2018ygx} where they took advantage of the Hofman-Maldacena map \cite{Hofman:2008ar} to move the light-sheet at infinity to a finite distance and they analized the structure of the algebra based on pure symmetry arguments. Therefore, it is of particular relevance to understand such algebra in the case of physical theories (possibly non-conformal) in 4 dimensions, which is the main purpose of this work. 

Here we interpret the insertion of light-ray operators in a correlation function as the insertion of a physical detector which is capturing properties of the distribution of outgoing particles at null infinity and we work in perturbation theory. In particular, we focus on the case of massless particles with integer spin and we use the saddle point estimate for the leading contribution to the stress tensor components. This provides an explicit representation of the hard mode contributions to light-ray operators, which is of particular relevance for physical detectors which have any kind of lower energy resolution. The structure we find is consistent with the quantum fields localizing on point-particle trajectory, since the stress tensor expression reduces to the pointlike one, possibly with the spin contribution  \cite{Pasterski:2015tva}. The universality of these expressions seems to point to some universality of light-ray operators in 4 dimensions beyond the conformal case. While a more rigorous analysis would be desired for the $[\mathcal{N}_{z}(\hat{\vec{n}}_1),\mathcal{N}_{\bar{z}}(\hat{\vec{n}}_2)]$ commutator, our results are consistent with the algebra of Cordova and Shao \cite{Cordova:2018ygx}. For the spin 1 case coupled with matter, an interaction term coming from the mixing of the so-called radiative and coulombic modes \cite{Ashtekar:2017wgq,Ashtekar:2017ydh} seems to break the algebra.

There could be extensions of the algebra in the soft sector for the angular momentum flux, in particular in the non-abelian case where we have observed a non-trivial boundary contribution whose insertion is related to the soft gluon theorem. This would required a detailed analysis of the Hilbert space factorization into soft and hard modes, which is beyond the scope of the paper  \cite{He:2020ifr}. More generally it would be interesting to establish a basis of non-local operators at null infinity, making contact also with integrals of current operators along the retarded time and to Wilson line operators. It would be also nice to have a full analysis of the algebra in the Bondi gauge, for example by using Newman-Penrose techniques  \cite{Newman:1961qr,Frolov:1977bp}. 

Borrowing some ideas from the QCD event shape detector operators, we have defined a suitable notion of energy event shape in the gravitational case using the Bondi news squared operator. Working with an EFT of gravity in the classical limit, we study the gravitational radiation produced in a particular direction $\hat{\vec{n}}$ in the sky from the scattering of compact objects in the leading soft approximation. In this simple limit, the Glauber-Sudarshan representation  \cite{Glauber:1963tx,Sudarshan:1963ts} of the QFT radiation density matrix reduces to a single coherent state and we can sum up all graviton contributions to get the classical result which matches the literature on the zero frequency limit (ZFL) \cite{Smarr:1977fy,Addazi:2019mjh}. Working in the approximation where we neglect the single graviton energies compared to the energy of the wave, we observe that there is a cancellation of infrared divergences for the gravitational energy event shape which gives an infrared finite result. It would be interesting to extend our analysis to include classical corrections to our result due to additional finite energy contributions and outside the zero frequency limit, as done in \cite{Gruzinov:2014moa,Addazi:2019mjh} for the soft gravitational radiation. In the classical limit, the correlation function of gravitational energy event shapes factorize because the graviton emissions are uncorrelated, which is in agreement with the single coherent state structure. In particular we have defined also the 2-point gravitational energy correlator, which is particularly interesting especially outside the soft classical limit where non-trivial contributions should come into play. Since it should be a non-trivial infrared safe prediction of quantum gravity, it would be very fascinating to explore this further. Interestingly, recently it was pointed out that a signature of the quantum state of gravitons might be observable especially if it deviates from the coherent state prediction  \cite{Parikh:2020nrd,Parikh:2020kfh,Parikh:2020fhy}. 

\nocite{Ashtekar:2014zsa}

\textbf{Acknowledgements}
We would like to thank Aron Wall, Simon Caron-Huot, Donal O'Connell, Manuel Accettulli-Huber and Tristan McLoughlin for useful discussions. We thank in particular Simon Caron-Huot for stimulating discussions at different stages of the project and for comments on the draft. R.G. would like to thank A.Cristofoli, D.Kosower and D.O'Connell for many discussions on a related project. This project has received funding from the European Union's Horizon 2020 research and innovation programme under the Marie Sk\l{}odowska-Curie grant agreement No. 764850. A.P. is grateful for funds provided by the Fonds de Recherche du Qu\'ebec.

\appendix

\section{Notations}
\label{sec:appendixA0}

The signature of the spacetime is mostly minus $(+,-,-,-)$. We have set $\hbar=c=1$.

\section{Detector light-cone definition of \texorpdfstring{$\mathcal{\tilde{E}}(\hat{\vec{n}})$}{En} and \texorpdfstring{$\mathcal{\tilde{K}}(\hat{\vec{n}})$}{Kn} light-ray operators}
\label{sec:appendixB}

To connect our discussion to the usual QCD event shape literature and in order to indroduce a more covariant formalism, we identify a detector by its set of coordinates $x^{\mu} = (t,r \hat{\vec{n}}) = \left(x_- n^{\mu} + x_+ \bar{n}^{\mu}\right) $ where
\begin{align}
x_+ &= \frac{t -r}{2} =  \frac{(x \cdot n)}{(n \cdot \bar{n})} \qquad x_- = \frac{t + r}{2} =  \frac{(x \cdot \bar{n})}{(n \cdot \bar{n})} ,\nonumber \\
n^{\mu} &= (1,\hat{\vec{n}}) \qquad \bar{n}^{\mu} = (1,-\hat{\vec{n}}) \qquad n^2 = \bar{n}^2 =0 .
\end{align}
We can complete the basis with two spacelike vectors\footnote{This is perfectly analogous to the Newman Penrose tetrad formalism.} $m_A$ which satisfy
\begin{align}
x_A &=  \frac{(x \cdot m_A)}{(m_A \cdot m_A)}  \qquad m_A \cdot n = m_A \cdot \bar{n} = 0 \qquad m_A \cdot m_B = - \delta_{A B} \qquad \forall \, A \in \{1,2\} ,
\end{align}
so that a generic point in spacetime is uniquely identified as $x^{\mu} = x_- n^{\mu} + x_+ \bar{n}^{\mu} + x_A m_A^{\mu}$. The completeness relation reads
\begin{align}
\eta_{\mu \nu} = \frac{n_{\mu} \bar{n}_{\nu} + \bar{n}_{\mu} n_{\nu}}{(n \cdot \bar{n})} + \frac{(m_A)_{\mu} (m_A)_{\nu}}{(m_A \cdot m_A)} ,
\end{align}
which implies that $x \cdot y = 2 (x_- y_+ + x_+ y_-) -  x_A y_A$ and that the flat metric can be written as $\mathrm{d} s^2 =  \mathrm{d} x^+ \mathrm{d} x^- -  \mathrm{d} x^A \mathrm{d} x^A$. It is worth pointing out that since we are interested in the algebra of different detectors it makes more sense to use a system of coordinates adapted at null infinity, as we did in the main text. Moreover, the latter extends to the more general case of asymptotically flat geometries and that is crucial for physical applications. 

We can write the ANEC operator at infinity in a covariant form as
\begin{align}
\mathcal{\tilde{E}}(\hat{\vec{n}}) =  (n \cdot \bar{n}) \int_{-\infty}^{+\infty} \mathrm{d} x_+ \, \lim_{x_- \to \infty} (x_-)^2 T_{--}(x_- n + x_+ \bar{n}) ,
\label{eqn:ANEC2}
\end{align}
where we have defined a covariant light-cone projection $T_{--} = \frac{\bar{n}^{\mu} \bar{n}^{\nu}}{(n \cdot \bar{n})^2} T_{\mu \nu}$. This expression is Lorentz covariant under rescalings of the type 
\begin{align}
n \to \rho \,n \qquad \bar{n} \to \rho \, \bar{n} 
\end{align}
and it has weight $-3$. It is important to stress that for external non-localized states, like momentum eigenstates, we need to take the large $x_-$ limit first and then perform the integral as done in \eqref{eqn:ANEC2}. The opposite is usually done in CFT for localized states, for which all notions are equivalent \cite{Hofman:2008ar,Belitsky:2013xxa}. In a similar way, we can write the momentum flow operator in a Lorentz covariant form as
\begin{align}
\mathcal{\tilde{P}}_{\mu}(\hat{\vec{n}}) =  (n \cdot \bar{n}) \int_{-\infty}^{+\infty} \mathrm{d} x_+ \, \lim_{x_- \to \infty} (x_-)^2 T_{-\mu}(x_- n + x_+ \bar{n}) .
\label{eqn:momentum_flow}
\end{align}
The next operator which we want to define is $\mathcal{\tilde{K}}(\hat{\vec{n}})$
\begin{align}
\mathcal{\tilde{K}}(\hat{\vec{n}}) &:=  (n \cdot \bar{n})^2 \int_{-\infty}^{+\infty} \mathrm{d} x_+ (x_+) \, \lim_{x_- \to \infty} (x_-)^2 T_{--}(x_- n + x_+ \bar{n}) ,
\end{align}
which has weight $-2$. The formulation of the operator $\mathcal{\tilde{N}}_A(\hat{\vec{n}})$, where the index $A$ has to be interpreted as a projection over the transverse direction $m_A$ ($\mathcal{\tilde{N}}_A(\hat{\vec{n}}) = \frac{(m_A)^{\mu} \mathcal{\tilde{N}}_{\mu}(\hat{\vec{n}})}{(m_A \cdot m_A)}$), is more difficult. This is related to the fact that one needs to rescale the index $A$ with $x_-$ to have a well-defined limit at infinity (see the related section \ref{sec:Cordova-Shao}).

\section{Light-ray operators for a light-sheet in the bulk}
\label{sec:appendixC}

In the general, one would like to generalize the light-ray operator definition to allow them to be localized on general light-sheets in the bulk of the spacetime. If the theory is scale invariant (i.e. for all CFT for example) then the two definitions will coincide (see the explicit Hofman-Maldacena conformal transformation  \cite{Hofman:2008ar}) but in general they will be different. To simplify our calculations, we can define
\begin{align}
\tilde{x}_{-} := \frac{x^0 + x^3}{2} \qquad \tilde{x}_+ := \frac{x^0 - x^3}{2} ,
\end{align}
so that light-sheets at fixed $\tilde{x}_- = \text{const}$ will be parametrized by the coordinate $\tilde{x}_+$. It is worth noticing that as $\tilde{x}_{-} \to +\infty$ we will have $\tilde{x}_{\pm} \to x_{\pm}$, so that there is no ambiguity with the previous notation in appendix B. For the light-sheet defined by $\tilde{x}_- = c$ with $c \in \mathbb{R}$, we define the light-ray family as
\begin{align}
\mathcal{E}(\tilde{x}_{\perp}) &:= \int_{-\infty}^{+\infty}  \mathrm{d} \tilde{x}_{+} \, T_{--}(\tilde{x}_+ = c,\tilde{x}_-,\tilde{x}_{\perp}) , \nonumber \\
\mathcal{K}(\tilde{x}_{\perp}) &:=  \int_{-\infty}^{+\infty} \mathrm{d} \tilde{x}_{+}\,(\tilde{x}_{+})\, T_{--}(\tilde{x}_+ = c,\tilde{x}_-,\tilde{x}_{\perp}) , \nonumber \\
\mathcal{N}_A(\tilde{x}_{\perp}) &:= (\partial_{A} \tilde{x}^{\mu}) \int_{-\infty}^{+\infty} \mathrm{d} \tilde{x}_{+} T_{- \mu}(\tilde{x}_+ = c,\tilde{x}_-,\tilde{x}_{\perp}) .
\label{eqn:def_E_NA_K}
\end{align}

\subsection{Spin 0}

For a general light-sheet $c \in \mathbb{R}$ one needs to compute 
\begin{align}
T_{--}(x) &=  \frac{1}{4} \left( T_{00}(x) + T_{33}(x) - 2 T_{03}(x) \right) =  \frac{1}{4} \left(\partial_0 \phi(x) - \partial_3 \phi(x)\right)^2 = \partial_- \phi(x) \partial_- \phi(x),\nonumber \\
T_{1-}(x) &= \frac{1}{2} \left( T_{10}(x) - T_{13}(x) \right) = \frac{1}{2}  \partial_1 \phi(x) \left(\partial_0 \phi(x) - \partial_3 \phi(x)\right) = \partial_1 \phi(x) \partial_- \phi(x) ,\nonumber \\
T_{2-}(x) &= \frac{1}{2} \left( T_{20}(x) - T_{23}(x) \right) = \frac{1}{2} \partial_2 \phi(x) \left(\partial_0 \phi(x) - \partial_3 \phi(x)\right) = \partial_2 \phi(x) \partial_- \phi(x) .
\end{align}
The light-ray operators are then defined as
\begin{align}
\mathcal{E}_{\text{scalar}}(\tilde{x}_{\perp}) &:=  \int_{-\infty}^{+\infty} \mathrm{d} \tilde{x}_{+} \, \partial_- \phi(\tilde{x}) \partial_- \phi(\tilde{x}) \Big|_{\tilde{x}_- = c} , \nonumber \\
\mathcal{N}_{A,\text{scalar}}(\tilde{x}_{\perp}) &:= \int_{-\infty}^{+\infty} \mathrm{d} \tilde{x}_{+} \, \partial_A \phi(\tilde{x}) \partial_- \phi(\tilde{x}) \Big|_{\tilde{x}_- = c} , \nonumber \\
\mathcal{K}_{\text{scalar}}(\tilde{x}_{\perp}) &:= \int_{-\infty}^{+\infty} \mathrm{d} \tilde{x}_{+} \, \tilde{x}_{+} \,\partial_- \phi(\tilde{x}) \partial_- \phi(\tilde{x}) \Big|_{\tilde{x}_- = c} .
\end{align}
It is quite remarkable that interactions drop out completely. This generically happens for scalar theories without derivative couplings and it implies some universality of the scalar light-ray algebra.

\subsection{Spin 1}

We have 
\begin{align}
T_{--} &= \frac{1}{2} \eta^{i j} \left[ (F_{0 i}^a - F_{3 i}^a) (F_{j 0}^b - F_{j 3}^b) \right] \text{Tr}(T^a T^b) = (\partial_-  A_1^a) (\partial_-  A_1^a) + (\partial_-  A_2^a) (\partial_-  A_2^a) ,\nonumber \\
T_{1-} &= F^{a}_{1 \beta} \eta^{\alpha \beta} (F_{\alpha 0}^b - F_{\alpha 3}^b) \text{Tr}(T^a T^b) =  (2 \partial_{[1} A_{2]}^a + g f^{a c d} A_1^c A_2^d) (\partial_{-} A_{2}^a) + 2 (\partial_{-} A^a_+) (\partial_{-} A^a_1) , \nonumber \\
T_{2-} &= F^{a}_{2 \beta} \eta^{\alpha \beta} (F_{\alpha 0}^b - F_{\alpha 3}^b) \text{Tr}(T^a T^b) = (2 \partial_{[2} A_{1]}^a + g f^{a c d} A_2^c A_{1}^d) (\partial_{-} A_{1}^a) + 2 (\partial_{-} A^a_+) (\partial_{-} A^a_2) ,
\end{align}
where in the last line have imposed the light-cone gauge
\begin{align}
A_0 = A_3 \Leftrightarrow A_- = 0 .
\end{align}
We see that while $T_{--}(\tilde{x})$ is interaction independent in this gauge, $T_{A-}(\tilde{x})$ is not. Nevertheless, we are interested in light-ray operators where we integrate over the light-cone time $\tilde{x}_+$ and therefore we can take advantage of the equations of motion
\begin{align}
\mathcal{D}_{\mu} F^{\mu,a}_{\,\,\,-} = \partial_- \left[\partial_1 A_1^a + \partial_2 A_2^a - 2 \partial_- A_+^a \right] + g f^{a b c} (A_1^b \partial_- A_1^c + A_2^b \partial_- A_2^c)  = 0 ,
\label{eqn:minus_eom}
\end{align}
provided that we take into account possible boundary terms at $\tilde{x}_+ \to \pm \infty$. To fix completely the gauge for the boundary contributions, we can enforce the radiation gauge where together with the light-cone condition $A_- = 0$ we have
\begin{align}
\nabla^{\mu} A_{\mu}^a(x)\Big|_{\mathcal{I}^-} = \lim_{\tilde{x}_+ \to -\infty} \left[2 \partial_- A_+^a - \partial_1 A_1^a - \partial_2 A_2^a\right] = 0 ,
\label{eqn:radiation_gauge_bulk}
\end{align}
thanks to which
\begin{align}
\int_{-\infty}^{+\infty} \mathrm{d} \tilde{x}_+  \, T_{1-}(\tilde{x}) &= \int_{-\infty}^{+\infty} \mathrm{d} \tilde{x}_+  \, \left[(\partial_{1} A_{2}^a - \partial_{2} A_{1}^a) (\partial_{-} A_{2}^a)+  (\partial_{1} A_{1}^a + \partial_{2} A_{2}^a) (\partial_{-} A_{1}^a) \right] \nonumber \\
&+ \frac{1}{2} A_1^a \left[2 \partial_- A_+^a - \partial_{1} A_{1}^a - \partial_{2} A_{2}^a \right]_{\tilde{x}_+ \to +\infty} , \nonumber \\
\int_{-\infty}^{+\infty} \mathrm{d} \tilde{x}_+  \, T_{2-}(\tilde{x}) &= \int_{-\infty}^{+\infty} \mathrm{d} \tilde{x}_+  \, \left[(\partial_{2} A_{1}^a - \partial_{1} A_{2}^a) (\partial_{-} A_{1}^a) + (\partial_{1} A_{1}^a + \partial_{2} A_{2}^a) (\partial_{-} A_{2}^a) \right] \nonumber \\
&+ \frac{1}{2} A_2^a \left[2 \partial_- A_+^a - \partial_{1} A_{1}^a - \partial_{2} A_{2}^a \right]_{\tilde{x}_+ \to +\infty} .
\end{align}
The light-ray operators are then defined as
\begin{align}
\mathcal{E}_{\text{gluon}}(\tilde{x}_{\perp}) &:= \sum_{i = 1}^2 \int_{-\infty}^{+\infty} \mathrm{d} \tilde{x}_+ \, \partial_- A_i^a(\tilde{x}) \partial_- A_i^a(\tilde{x}) \Big|_{\tilde{x}_- = c} , \nonumber \\
\mathcal{K}_{\text{gluon}}(\tilde{x}_{\perp}) &:=\sum_{i = 1}^2 \int_{-\infty}^{+\infty} \mathrm{d} \tilde{x}_+ \, \tilde{x}_{+} \, \partial_- A_i^a(\tilde{x}) \partial_- A_i^a(\tilde{x}) \Big|_{\tilde{x}_- = c} , \nonumber \\
\mathcal{N}_{1,\text{gluon}}(\tilde{x}_{\perp}) &:= \int_{-\infty}^{+\infty} \mathrm{d} \tilde{x}_+ \, \left[(\partial_{1} A_{2}^a - \partial_{2} A_{1}^a) (\partial_{-} A_{2}^a)+  (\partial_{1} A_{1}^a + \partial_{2} A_{2}^a) (\partial_{-} A_{1}^a) \right] \Big|_{\tilde{x}_- = c} \nonumber \\
&+\frac{1}{2} A_1^a \left[2 \partial_- A_+^a - \partial_{1} A_{1}^a - \partial_{2} A_{2}^a \right]_{\tilde{x}_+ \to +\infty} ,\nonumber \\
\mathcal{N}_{2,\text{gluon}}(\tilde{x}_{\perp}) &:= \int_{-\infty}^{+\infty} \mathrm{d} \tilde{x}_+ \,  \left[(\partial_{2} A_{1}^a - \partial_{1} A_{2}^a) (\partial_{-} A_{1}^a) + (\partial_{1} A_{1}^a + \partial_{2} A_{2}^a) (\partial_{-} A_{2}^a) \right]  \Big|_{\tilde{x}_- = c} \nonumber \\
&+\frac{1}{2} A_2^a \left[2 \partial_- A_+^a - \partial_{1} A_{1}^a - \partial_{2} A_{2}^a \right]_{\tilde{x}_+ \to +\infty} .
\label{eqn:Lightrayspin1_finitec}
\end{align}
If we compare to the self-interacting scalar case, we see that in YM theory interactions do not affect the definition of the operators $\mathcal{E}_{\text{gluon}}(\tilde{x}_{\perp})$ and $\mathcal{K}_{\text{gluon}}(\tilde{x}_{\perp})$ but they appear in $\mathcal{N}_{A,\text{gluon}}(\tilde{x}_{\perp})$ through the boundary contribution. 

\section{Shear-inclusive ANEC and Raychaudhuri equation}
\label{sec:appendixD}

The connection of the ANEC operator and its expectation value with causality properties of the underlying quantum field theory is well-known  \cite{Hartman:2016lgu}. A congruence of null (complete, achronal) geodesics obeys Raychaudhuri equation
\begin{align}
\frac{\mathrm{d} \Theta}{\mathrm{d} \lambda} = - \frac{1}{2} \Theta^2 - \sigma_{\mu \nu} \sigma^{\mu \nu} - 8 \pi G T^{\text{matter}}_{\mu \nu} k^{\mu} k^{\nu} ,
\label{eqn:Raychaudhuri}
\end{align}
where $\lambda$ is the affine parameter, $\Theta = \frac{1}{A} \frac{\mathrm{d} A}{\mathrm{d} \lambda}$\footnote{$A$ is the cross sectional area here.} is the expansion,  $\sigma_{\mu \nu}$ is the shear and $k^{\mu}$ is a null vector of unit affine length which in our case we will take the light-sheet generator $k^{\mu} = \delta^{\mu}_-$. The right hand side of \eqref{eqn:Raychaudhuri} immediately reminds us the enlarged definition of the matter ANEC with the gravitational contribution once we integrate along the null direction of the generator, albeit here \eqref{eqn:Raychaudhuri} is purely classical in its nature. If we consider a generic congruence of null geodesics with starts at $\mathcal{I}^{-}$ and ends on $\mathcal{I}^{+}$, one can ask how the expansion parameter changes along the congruence. Classically, the ANEC condition $T^{\text{matter}}_{\mu \nu} k^{\mu} k^{\nu} \geq 0$ implies the classical focussing theorem, that is light-rays never ``anti-focus'' as long as matter has positive energy. Quantum mechanically, quantum fields can developed negative energy so that we generically need a quantum version of the ANEC  \cite{Faulkner:2016mzt} and of the focussing theorem  \cite{Bousso:2015mna}. In general, there are problems in including quantized gravity perturbations in this story.\footnote{We thank Aron Wall for useful discussions on this point.} Nevertheless, assuming we have matter minimally coupled with gravity, for a stationary null surface one can quantize $h_{\mu \nu}$ on the null surface in light-front quantization  \cite{Wall:2010cj} and at the first order in the perturbation one can define a shear-inclusive ANEC operator of the form
\begin{align}
\mathcal{E}_{\text{shear-inclusive}} = \int \mathrm{d} \lambda  \left[\frac{1}{8 \pi G}\sigma_{\mu \nu} \sigma^{\mu \nu} + T^{\text{matter}}_{\mu \nu} k^{\mu} k^{\nu} \right] ,
\end{align}
whose expectation value is positive definite $\langle \mathcal{E}^{\text{shear-inclusive}} \rangle \geq 0$ as it can be proved from the linearized Raychaudhuri equation  \cite{Wall:2009wi}. The connection to our previous definition of the shear-inclusive ANEC in the Bondi gauge consists in considering the asymptotically flat region for the geodesic congruence. Indeed in such region one can define a boundary shear as the limit of the rescaled shear of the congruence on future null infinity so that
\begin{align}
\lim_{r \to +\infty} r^2 \left[\frac{1}{8 \pi G} \,\sigma_{\mu \nu} \sigma^{\mu \nu} + T^{\text{matter}}_{\mu \nu} k^{\mu} k^{\nu} \right] \sim \frac{1}{32 \pi G} N_{\zeta \zeta} N^{\zeta \zeta} + \lim_{r \to \infty} r^2 T^{\text{matter}}_{v v} ,
\end{align}
as proved in \cite{Bousso:2016vlt,Bousso:2016wwu} in order to study asymptotic entropy bounds.







\bibliographystyle{JHEP-2}
\bibliography{Lightray}

\end{document}